\documentclass[11pt]{article}
\pdfoutput=1

\usepackage{xcolor}
\definecolor{darkblue}{RGB}{0,0,139}
\definecolor{darkgreen}{RGB}{0,100,0}
\definecolor{reddish}{RGB}{180,0,0}

\usepackage{hyperref}
\hypersetup{,
	pdftitle={MSCSs},
  	pdfauthor={Rybak, Maji, Shafi},
    colorlinks=true,       
    linkcolor=darkblue,          
    citecolor=darkgreen,        
    filecolor=reddish,      
    urlcolor=blue,           
    linktocpage=true
}

\usepackage{euscript}
\usepackage{amsfonts}
\usepackage{amsbsy}
\usepackage{epsfig}
\usepackage{amsthm}
\usepackage{amscd}
\usepackage{amstext}
\usepackage{verbatim}
\usepackage{cancel}
\usepackage{capt-of}
\usepackage{empheq}

\usepackage[nosort]{cite}
\usepackage{bm}
\usepackage{authblk}
\usepackage{esint}
\usepackage[T1]{fontenc}
\usepackage{mathdots}
\usepackage[centertableaux]{ytableau}

\usepackage[utf8]{inputenc}
\usepackage{graphicx}
\usepackage{physics}
\usepackage{mathtools}
\usepackage{bbold}

\usepackage{setspace}
\usepackage{colortbl}
\usepackage{xcolor}

\usepackage{amsmath}
\makeatletter
\renewcommand*\env@matrix[1][\arraystretch]{%
  \edef\arraystretch{#1}%
  \hskip -\arraycolsep
  \let\@ifnextchar\new@ifnextchar
  \array{*\c@MaxMatrixCols c}}
\makeatother
\usepackage{amssymb}
\usepackage{mathrsfs}
\usepackage{booktabs}
\usepackage{bbm}
\usepackage{float}

\usepackage{pgfplots}
\pgfplotsset{compat=1.15}
\usepackage{tikz}
\usetikzlibrary{external}
\usetikzlibrary{arrows}

\usepackage{subcaption}
\usepackage{graphicx}
\usepackage{mathtools}
\usepackage[many]{tcolorbox}
\tcbset{shield externalize}
\usepackage{xcolor, colortbl}
\usepackage[a4paper]{geometry}
\usepackage[normalem]{ulem}

\usepackage{mathrsfs}

\title{Gravitational Wave Background from Metastable Current-Carrying Cosmic Strings}

\author[a]{Ivan Rybak}
\author[a]{Rinku Maji}
\author[b]{Qaisar Shafi}
\affil[a]{\textit{Cosmology, Gravity, and Astroparticle Physics Group\\
Center for Theoretical Physics of the Universe,
Institute for Basic Science (IBS)\\
Daejeon 34126, Korea}}
\affil[b]{\textit{Bartol Research Institute, Department of Physics and Astronomy\\
University of Delaware, Newark, DE 19716, USA}}

\begin{document}

\maketitle

\begin{abstract}
We study the stochastic gravitational wave background (SGWB) generated by networks of current-carrying metastable cosmic strings (MSCSs). Extending previous studies, we include the possibility of vorton formation phenomenologically.
We construct an analytical template that provides a fast and accurate approximation to the spectrum in the frequency range probed by pulsar timing arrays (PTAs). We also show how the parameters of cosmic string networks determine the low-frequency slope. For the frequency range relevant for LIGO-Virgo-KAGRA (LVK), we derive an analytical expression that constrains the scale-invariant part of the SGWB. Finally, we confront the model with current constraints from PTAs and the LVK collaboration. We identify regions of the parameter space that can produce a signal consistent with PTA observations but are in tension with LVK constraints, suggesting that additional mechanisms to suppress the high-frequency spectrum are required for a consistent interpretation of the PTA signal.
\end{abstract}

\tableofcontents

\section{Introduction}
\label{introduction}

The early Universe, characterized by a particular symmetry-breaking scenario, can give rise to the formation of one-dimensional topological defects \cite{Kibble:1976sj, Kibble:1999yk}; for reviews, see \cite{Hindmarsh:1994re,Vilenkin:2000jqa}. These one-dimensional topological defects, known as cosmic strings, are predicted in a wide range of theories beyond the Standard Model (BSM) \cite{Lazarides:1981fv, Kibble:1982ae, Lazarides:1982tw, Davis:1986xc, Brandenberger:2013xyk, Afzal:2023cyp, Maji:2025thf}. Studying the observational signatures of cosmic strings allows us to constrain their properties, thereby probing the corresponding high-energy models of the early Universe.

One of the most powerful observational probes of cosmic strings is the cosmic microwave background (CMB) \cite{Battye:2006pk, Lazanu:2014eya, Lizarraga:2016onn, Ringeval:2015ywa, Rybak:2021scp, Raidal:2026cpb}. However, even greater sensitivity may be achieved through the detection of gravitational waves emitted by cosmic string loops \cite{Caldwell:1996en, Sousa:2013aaa, Blanco-Pillado:2017oxo}.\footnote{Most predictions for the gravitational wave signal rely on the infinitely thin string approximation, under which gravitational wave emission is the dominant energy-loss mechanism for cosmic string loops. In contrast, field-theory simulations predict that loops lose energy mainly through the emission of massive particles \cite{Hindmarsh:2021mnl, Baeza-Ballesteros:2024sny}.} Gravitational wave observations provide a powerful method of probing a broad class of high-energy physics models \cite{Cui:2017ufi, Gouttenoire:2019kij}. Cosmic string scenarios have attracted even greater interest due to results from pulsar timing array (PTA) observational data \cite{NANOGrav:2023gor, Reardon:2023gzh, EPTA:2023xxk}, which provide evidence for a stochastic gravitational wave background (SGWB). The observed signal can be interpreted within the framework of metastable cosmic string (MSCS) models, motivating further studies of their gravitational wave signatures \cite{Buchmuller:2020lbh, NANOGrav:2023hvm, Buchmuller:2023aus, Maji:2022jzu, Ahmad:2025dds}. 

Most studies describe cosmic strings based on the Abelian–Higgs theory or its infinitely thin approximation, given by the Nambu–Goto (NG) action. Within these descriptions, any additional fields whose excitations may propagate along the string core are typically neglected. However, in many realistic scenarios, this is not the case, and we cannot neglect the presence of other fields that couple to the string core \cite{Peter:1993tm, Davis:1996xs, Davis:1997bs, Allys:2015yda, Fukuda:2020kym, Maji:2025itv}. One of the earliest examples of a cosmic string model incorporating additional degrees of freedom was proposed in \cite{Witten:1984eb}. In addition, these propagating degrees of freedom may also emerge within an effective description of string dynamics \cite{Carter:1994zs, Carter:2000wv, Vieira:2016vht}.

Realistic cosmic string scenarios with many interacting fields \cite{Allys:2015kge, Abe:2020ure, Lazarides:2022ezc, Fu:2023mdu} motivate us to go beyond the standard NG approximation and incorporate worldsheet currents into the description of cosmic strings. In this work, we focus on the phenomenology of current-carrying MSCSs and investigate their gravitational wave signatures in light of PTA observations \cite{NANOGrav:2023gor}.

Field-theory simulations can provide a first-principles description of current-carrying cosmic strings, and significant progress in this direction has been achieved in recent years \cite{Hiramatsu:2023epr, Correia:2024wsq, Jhun:2026iky}. However, as with Abelian-Higgs string simulations, current simulations cannot resolve a realistic cosmological setup, and we therefore need to rely on extrapolations over many orders of magnitude. In our approach, we use the infinitely thin approximation, which includes additional degrees of freedom that propagate along the string core. We also assume that long strings decay into loops that predominantly decay into massless radiation, as in \cite{Blanco-Pillado:2017oxo, Lorenz:2010sm, Auclair:2019wcv}. In particular, we describe the long-string network using the effective charge-velocity-dependent one-scale (CVOS) model \cite{Rybak:2017yfu, Martins:2020jbq, Martins:2021cid}. However, we recognize that ongoing debates over the dynamics of NG and Abelian-Higgs cosmic string loops reflect substantially different predictions for gravitational wave radiation \cite{Hindmarsh:2021mnl, Baeza-Ballesteros:2024sny}, and we parameterize this uncertainty.

The remainder of this paper is organized as follows. In Section \ref{II}, we review the evolution of current-carrying MSCS networks and derive the corresponding loop distribution function. In Section \ref{III}, we investigate the SGWB generated by these loops and identify regions of the parameter space characterized by different low-frequency spectral slopes. In Section \ref{IV}, we develop a computationally efficient approximation for the SGWB spectrum that is suitable for exploring the cosmic string parameter space with PTA and LVK. Finally, in Section \ref{V}, we analyze PTA data using PTArcade \cite{Mitridate:2023oar} and incorporate constraints on a scale-invariant SGWB from the LVK collaboration \cite{LIGOScientific:2025bgj}. We conclude in Section \ref{VI}.

 \section{Evolution of metastable current-carrying cosmic strings}
 \label{II}

\subsection{Description of individual current-carrying strings}

We begin by considering the action for a generic current-carrying cosmic string. For a detailed review, see Ref.~\cite{Carter:1997pb}. The action is given by
\begin{equation}
\label{Action}
S = \mu \int \mathcal{L}(\chi) \sqrt{-\gamma} d^2\zeta ,
\end{equation}
where $\mu$ sets the energy scale of the symmetry-breaking phase transition responsible for cosmic string formation. The function $\mathcal{L}(\chi)$ is the master function that characterizes the string equation of state and depends on the scalar invariant
\begin{equation}
\chi \equiv \gamma^{ab} \varphi_{,a} \varphi_{,b},
\end{equation}
where $\varphi(\zeta^a)$ is a scalar field living on the two-dimensional string worldsheet. The induced worldsheet metric is
\begin{equation}
\gamma_{ab} = X^\mu_{,a} X^\nu_{,b} g_{\mu\nu},
\end{equation}
with determinant $\gamma$. Here, $X^\mu(\zeta^a)$ defines the embedding of the string worldsheet into the four-dimensional spacetime with metric $g_{\mu\nu}$. The worldsheet coordinates are denoted by $\zeta^a$ ($a,b=0,1$), while spacetime indices are $\mu,\nu=0,\ldots,3$. Throughout this work, a comma denotes differentiation with respect to the corresponding coordinate, e.g.,
\begin{equation}
X^\mu_{,a}\equiv\frac{\partial X^\mu}{\partial\zeta^a}.
\end{equation}

We consider the evolution of current-carrying cosmic strings in a spatially flat Friedmann-Lemaître-Robertson-Walker (FLRW) universe with line element
\begin{equation}
ds^2=a^2(\tau)\left(d\tau^2-d\mathbf{x}^2\right),
\end{equation}
where $a(\tau)$ is the scale factor and $\tau$ denotes conformal time.

Throughout this paper, we work in the temporal-transverse gauge, in which the worldsheet coordinates are chosen as $\zeta^a=(\tau,\sigma)$. The gauge conditions are
\begin{equation}
X^0=\tau,\qquad
\dot{\mathbf{X}}\cdot\mathbf{X}'=0,
\end{equation}
where $\dot{\mathbf{X}}\equiv\frac{\partial\mathbf{X}}{\partial\tau},$ $\mathbf{X}'\equiv\frac{\partial\mathbf{X}}{\partial\sigma}.$

In this study, we will consider chiral currents, i.e., the case in which the current is described by a light-like gradient, $\chi \rightarrow 0$. This is the simplest model that nevertheless corresponds to physically relevant realizations of current-carrying strings \cite{Carter:1999hx, Carter:2003fb} and appears to be the most relevant case for describing cosmic string networks \cite{Rybak:2023jjn}. A particular property of chiral strings is that their string tension and mass per unit length are the same as those of ordinary cosmic strings. Thus, we can anticipate that the standard expression for the semiclassical decay rate remains valid \cite{Preskill:1992ck, Vilenkin:1982hm}. Hence, we use the following expression:
\begin{equation}
\label{Probability}
\Gamma_d = \frac{\mu }{2 \pi}
\exp\left[- \pi \kappa_d \right],
\end{equation}
where $\kappa_d \equiv \frac{m^2}{\mu}$ and $m$ is the monopole mass. A detailed derivation can be found in Appendix \ref{Monopole-antimonopole nucleation for current-carrying cosmic strings}.

 \subsection{Evolution of the metastable current-carrying cosmic string network}

The evolution of a standard cosmic string network formed during an early-universe symmetry-breaking phase transition approaches a scaling regime characterized by
\begin{equation}
\label{Scaling}
v=\mathrm{const.}, \qquad
\rho=\frac{\mathrm{const.}}{t^2},
\end{equation}
where the root-mean-square (RMS) velocity $v$ approaches a constant value and the string energy density decreases inversely with the square of the physical time $t = \int a d\tau$. This behavior was first proposed in Ref.~\cite{Kibble:1984hp} and has since been confirmed by numerous NG simulations \cite{Bennett:1987vf,Bennett:1989yp,Bennett:1990uza,Ringeval:2005kr,Martins:2005es,Blanco-Pillado:2011egf}, field-theory simulations \cite{Hindmarsh:2017qff,Correia:2020yqg,Correia:2021tok}, and semi-analytical models \cite{Bennett:1985qt,Bennett:1986zn,Austin:1993rg,Martins:1996jp,Martins:2000cs}. Semi-analytical studies have further demonstrated that current-carrying cosmic string networks also evolve toward scaling solutions during the radiation-dominated epoch, although the corresponding scaling values generally differ from those of ordinary cosmic strings \cite{Martins:2021cid,Rybak:2023jjn}.

During the scaling regime, the string energy density evolves as \eqref{Scaling}. Since the volume of the particle horizon scales as $V_h\propto t^3$, the total energy stored in the long strings within the horizon is $\rho V_h$. In the chiral limit considered in this work, the energy per unit length is simply $\mu$, and therefore the total length of strings contained within the horizon is $L_h=\frac{\rho V_h}{\mu}$.
Assuming that the decay rate per unit string length and per unit time is $\Gamma_d$, the expected probability of monopole-antimonopole nucleation events occurring within one Hubble time is given by $d P \simeq \Gamma_d L_h dt$, and the number of decay events per unit Hubble volume is then given by
\begin{equation}
N_d = \int d P \simeq \int \Gamma_d L_h dt \simeq \Gamma_d t^2.
\end{equation}
The effects of metastability become significant when the expected number of decay events within the horizon is of the order of unity, i.e., when $N_d\sim1$. This defines the characteristic timescale
\begin{equation}
\label{t_s}
t_s\simeq\Gamma_d^{-1/2},
\end{equation}
before which the effects of string decay are expected to be negligible, and the evolution of an MSCS network is the same as for a stable network, while for $t>t_s$ the string network experiences rapid decay.

To describe the evolution of current-carrying cosmic string networks, we use the CVOS model. As discussed above, we restrict our analysis to the chiral current, whose macroscopic evolution is given by the following system of equations \cite{Rybak:2023jjn}
\begin{subequations}
\label{EqOfMotMacroLin2}
\begin{align}
\label{EqOfMotMacroLinA2}
\frac{d\xi}{dt}
&= \xi H \left( 1+\frac{v^2}{1+Y} \right) +\frac{cv}{2} +\frac{Yvk(v)}{1+Y}, \\
\label{EqOfMotMacroLinB2}
\frac{dv}{dt} &= \frac{1-v^2}{1+Y} \left[ \frac{(1-Y)k(v)}{\xi} -2Hv \right],\\
\label{EqOfMotMacroLinC2}
\frac{dY}{dt} &= 2Y \left( \frac{vk(v)}{\xi} -H \right) - A(Y)\frac{Y}{\xi},
\end{align}
\end{subequations}
where the Hubble parameter is defined by $ H\equiv\frac{1}{a}\frac{da}{dt}$, and the macroscopic variables describing the string network are:
\begin{itemize}
\item $\xi$, the characteristic length of the network, related to the bare string energy density through $\rho_0=\mu/\xi^2$;
\item $v$, the RMS velocity of the strings;
\item $Y$, the average dimensionless charge amplitude of the network.
\end{itemize}

A scaling solution corresponds to $v=\mathrm{const.}$, $Y=\mathrm{const.}$, and $\xi=\zeta t$, where $\zeta$ is also a constant. Different scaling solutions of the CVOS model were recently investigated in Refs.~\cite{Pimenta:2024blk,Pimenta:2025aak}. Throughout this work, we adopt the same momentum parameter and loop-chopping efficiency as those used for NG strings \cite{Martins:2000cs}: $c=0.23$ and 
\begin{equation}
\label{Momentum}
k(v)= \frac{2\sqrt{2}}{\pi} \frac{1-8v^6}{1+8v^6} (1-v^2) (1+2\sqrt{2}v^3).
\end{equation}
The function $A(Y)$ represents the charge leakage function, whose form depends on the underlying microscopic model. Its influence on the evolution of current-carrying cosmic string networks has been investigated in detail in Refs.~\cite{Martins:2021cid,Rybak:2023jjn}.

We consider the evolution of the string network in the expanding universe described by the Friedmann equation,
\begin{equation}
H^2= H_0^2 \left[ \Omega_r \left( \frac{a_0}{a} \right)^4 + \Omega_m \left( \frac{a_0}{a} \right)^3 + \Omega_\Lambda \right],
\end{equation}
where we use the following cosmological parameters: $\Omega_m=0.308$, $\Omega_r = 9.1476 \times 10^{-5}$, and $\Omega_{\Lambda}=1-\Omega_m-\Omega_r$ from Ref.~\cite{Planck:2018vyg}.

\begin{figure}[t!]
\begin{center}
\includegraphics[scale=0.45]{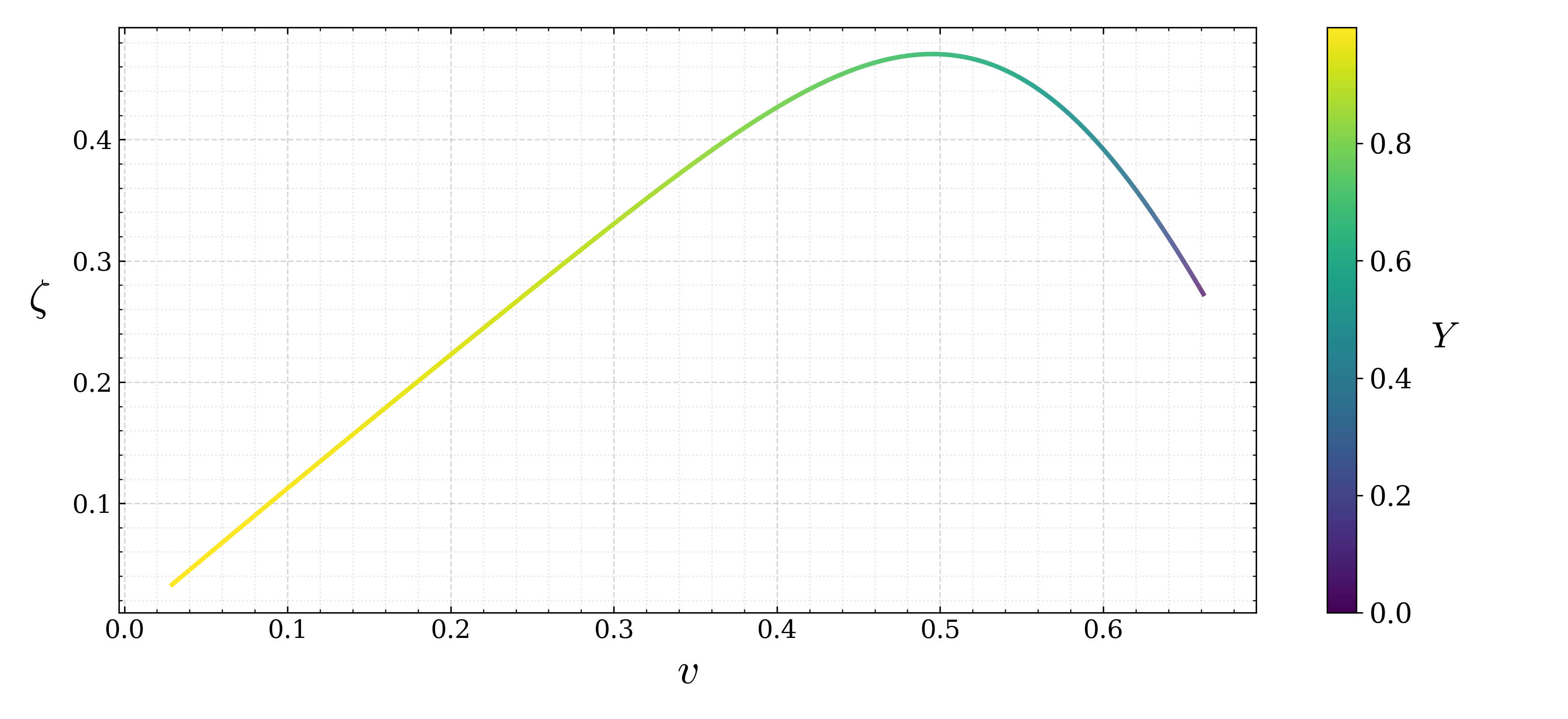}
\caption{\label{Fig.Zeta_V_Y}
Scaling values of the characteristic length parameter $\zeta$ and the RMS velocity $v$ as functions of the charge amplitude $Y$, obtained from the stationary solutions of Eq.~\eqref{AlgebrEq}.}
\end{center}
\end{figure}

Since the charge leakage function $A(Y)$ is model-dependent and our primary interest is the effect of the current during the radiation-dominated epoch, when the string network has already reached the scaling regime, we treat the current amplitude $Y$ as a free parameter. This allows us to keep the analysis generic and independent of a particular microscopic realization of the current-carrying string model. Hence, we can reduce the system of differential equations~\eqref{EqOfMotMacroLin2} to the algebraic equations determining the stationary scaling solution for $\zeta$ and $v$ as functions of $Y$ (a detailed stability analysis can be found in Ref.~\cite{Martins:2021cid}),
\begin{equation}
\label{AlgebrEq}
\zeta = \frac{ c v (1+Y) + 2 Y v k(v)}{1+Y-v^2},
\qquad v = \frac{(1-Y)k(v)}{\zeta},
\end{equation}
where we have excluded the $v=1$ case.
The resulting dependence of the scaling variables on the charge amplitude $Y$ is shown in Fig.~\ref{Fig.Zeta_V_Y}. To enable efficient computation of the SGWB template, we use parametrized interpolations for the functions $\zeta(Y)$ and $v(Y)$.

\subsection{Loop distribution function}

In this subsection, we establish the form of the loop distribution function for current-carrying MSCSs. We assume that the string network evolves in the scaling regime and that all loops are produced with the same initial size, which is fixed by a fraction of the characteristic network length,
\begin{equation}
l = \alpha \xi.
\end{equation}
This assumption was verified in NG simulations of ordinary cosmic strings \cite{Blanco-Pillado:2011egf,Blanco-Pillado:2013qja,Blanco-Pillado:2019tbi}. However, no analogous simulations have been performed for current-carrying cosmic strings, and we therefore assume that this relation remains valid for current-carrying strings.

The CVOS model, given by Eqs.~\eqref{EqOfMotMacroLin2}, describes the evolution of the long-string network, whose characteristic length is of the order of the cosmological horizon. The loop chopping efficiency parameter $c$ in Eq.~\eqref{EqOfMotMacroLinA2} describes the transfer of energy from the long-string network into closed loops through intercommutations. Once formed, these loops oscillate and emit gravitational radiation. For standard NG cosmic strings, they are the dominant sources of the SGWB, provided that the initial loop size satisfies $\alpha > \Gamma G\mu$, where $\Gamma$ is the gravitational wave emission efficiency parameter for loops \cite{Allen:1991bk}.

The loop distribution function $n(l(t),t)$ for standard MSCSs was derived in \cite{Buchmuller:2020lbh,Leblond:2009fq}. Since current-carrying cosmic strings have a loop-production mechanism similar to that of standard cosmic strings, we expect that an analogous loop distribution function should be valid for current-carrying cosmic strings as well:
\begin{equation}
    \label{LoopNumbDensG2}
    n(l(t),t) = \mathcal{F} \sqrt{1-v_b^2} \frac{c v_b}{\alpha \xi_b^4} \left( \frac{a_b}{a(t)} \right)^3   \frac{ \exp \left[- \Gamma_d \left( l ( t - t_b ) + \Gamma_{\rm tot} \frac{\left( t - t_b \right)^2}{2}  \right) \right] }{ \alpha \frac{d \xi}{dt} \big|_{t=t_{b}} + \Gamma_{\rm tot} } ,
\end{equation}
A detailed derivation is given in Appendix \ref{Loop distribution function for metastable current-carrying cosmic string network}. Here, $\mathcal{F}$ is a fuzziness parameter that determines the fraction of loops whose energy is emitted as gravitational or electromagnetic massless radiation. From NG simulations \cite{Blanco-Pillado:2013qja}, it was established that $\mathcal{F} \approx 0.1$, accounting for small loops whose energy is redshifted due to the expansion of the Universe and which do not contribute to the SGWB. By changing $\mathcal{F}$, we can effectively account for the effect of some fraction of loops decaying via massive-particle radiation, thereby parameterizing the uncertainty between NG and field-theory simulations. The subscript $b$ in Eq.~\eqref{LoopNumbDensG2} indicates that these quantities are evaluated at the loop formation time $t_b$. We also use the fact that the length of a string loop decreases through the emission of gravitational and vector radiation as \cite{Rybak:2024our}\footnote{Small loops may also radiate massive particles, which can modify the SGWB prediction \cite{Matsunami:2019fss,Auclair:2019jip,Rybak:2023bky}. In this work, we neglect this contribution.}
\begin{equation}
\label{LengthChange}
    l(t) = l(t') - \Gamma_{\rm tot} \left( t - t' \right),
\end{equation}
with $\Gamma_{\rm tot} \equiv \Gamma(Y) G \mu + e^2 \Gamma^V(Y)$, and $ \Gamma(Y) = 50 \left( 1 - Y^{1/2} \right)^2$,  $\Gamma^V(Y) = 9 Y^{1/2} \left( 1 - Y^{1/2} \right)$ according to \cite{Rybak:2024our,Rybak:2022sbo}, where $e^2$ is the coupling between the current carriers and the vector field. We further assume that the current parameter $Y_L$ remains constant throughout the lifetime of a loop. Thus, $\Gamma$ and $\Gamma^V$ are treated as constants, although, in general, they evolve with the loop current amplitude $Y_L$ \cite{Rybak:2024our,Rybak:2022sbo}.

The expression~\eqref{LoopNumbDensG2} reproduces the result obtained in \cite{Buchmuller:2021mbb} for standard cosmic strings in the limit of vanishing current. In contrast to Ref.~\cite{Buchmuller:2021mbb}, however, we will explicitly use Eq.~\eqref{LoopNumbDensG2}, numerically determining the loop formation time $t_b$ and substituting it directly into the SGWB calculation. This approach eliminates the need for the approximation $\alpha \zeta/\Gamma_{\rm tot} \gg 1$, extending the analysis to a wider region of parameter space, including the regime of small loops or large $\Gamma_{\rm tot}$.

\section{Numerical results for the SGWB}
\label{III}

\subsection{General formalism}

In this section, we provide details of the full numerical calculations for the SGWB from current-carrying MSCSs. The SGWB spectrum can be written as
\begin{equation}
\label{OmegaSGWB}
    \Omega_{\rm gw} (f) = \sum_{j=1}^{N}  \Omega^j_{\rm gw}, 
\end{equation}
where the sum is truncated at $N$ due to string thickness and gravitational backreaction, as discussed in \cite{Caldwell:1996en, Sanidas:2012ee, Drew:2025txu, Gerblich:2026boa}\footnote{The formalism and the influence of gravitational back-reaction on SGWB produced by standard Nambu-Goto string loops can be found in Refs.~\cite{Chernoff:2018evo, Wachter:2024aos, Wachter:2024zly }.}. The contribution from the $j$th harmonic is given by 
\begin{equation}
\label{Omega_j}
   \Omega^j_{\rm gw} =   \frac{16 \pi}{3} \left( \frac{G \mu}{H_0} \right)^2 \frac{2 j}{f} \int_{t_i}^{t_0} P_j n(l,t) \left( \frac{a(t)}{a_0} \right)^5 dt,
\end{equation}
where $t_i$ is the time at which the string network is formed, and the power spectrum is given by \cite{Rybak:2024our}
\begin{equation}
    P_j = \Gamma_0 j^{-q} \text{e}^{- j g(Y)} \frac{\left( 1 - Y^{1/2} \right)^B}{\mathcal{E}(Y)} ,
\end{equation}
with $\mathcal{E}(Y) = \sum_{j=1}^{N} j^{-q} \text{e}^{-j g(Y)} $, $B=2$, $\Gamma_0 = 50$ and
\begin{equation}
\label{gCK}
g(\mathcal{Y}) \approx \begin{cases} 8 \left( 1 - \left( 1-\mathcal{Y} \right)^{1/4} \right)^{2} , & \quad q=4/3 \quad \text{(quasi-cusps)} \\
0 \;, \qquad & \quad q=5/3 \quad \text{(kinks)}
\end{cases}
\end{equation}

The relation between string loop length and the frequency of radiation is given by $l= \frac{2 j}{f} \frac{a(t)}{a_0} S(Y)$ with $S(Y) = \frac{2 \sqrt{1-Y}}{1+\sqrt{1-Y}}$ \cite{Rybak:2024our}.
Using the numerical solution for $t_b$ that is given by the equation
\begin{equation}
    2 \frac{j}{f} \frac{a(t)}{a_0} S(Y) = \alpha \xi(t_b) - \Gamma_{\rm tot} \left( t - t_b \right)
\end{equation}
we substitute it into Eq.~\eqref{LoopNumbDensG2} to express the loop distribution function $n(l,t)$. Together with expressions for cosmic string network evolution, given by Eq.~\eqref{EqOfMotMacroLin2}, we can numerically obtain the SGWB for current-carrying MSCS networks that are not necessarily in the scaling regime, as was previously studied for NG strings in Ref.~\cite{Sousa:2013aaa}. We also include the change in the number of relativistic degrees of freedom in the Standard Model (SM) by making the substitution 
\begin{equation}
\Omega_r \rightarrow \Omega_r \Delta_r(a) \, , \text{with} \qquad \Delta_r(a) = \frac{g_*(a)}{g_*(a_0)} \left( \frac{g_{*S}(a_0)}{g_{*S}(a)} \right)^{4/3},    
\end{equation}
where the corresponding factors are interpolated with the use of \texttt{micrOMEGAs5.0} \cite{Belanger:2018ccd}.  

\begin{figure}[t!]
\begin{center}
\includegraphics[scale=0.65]{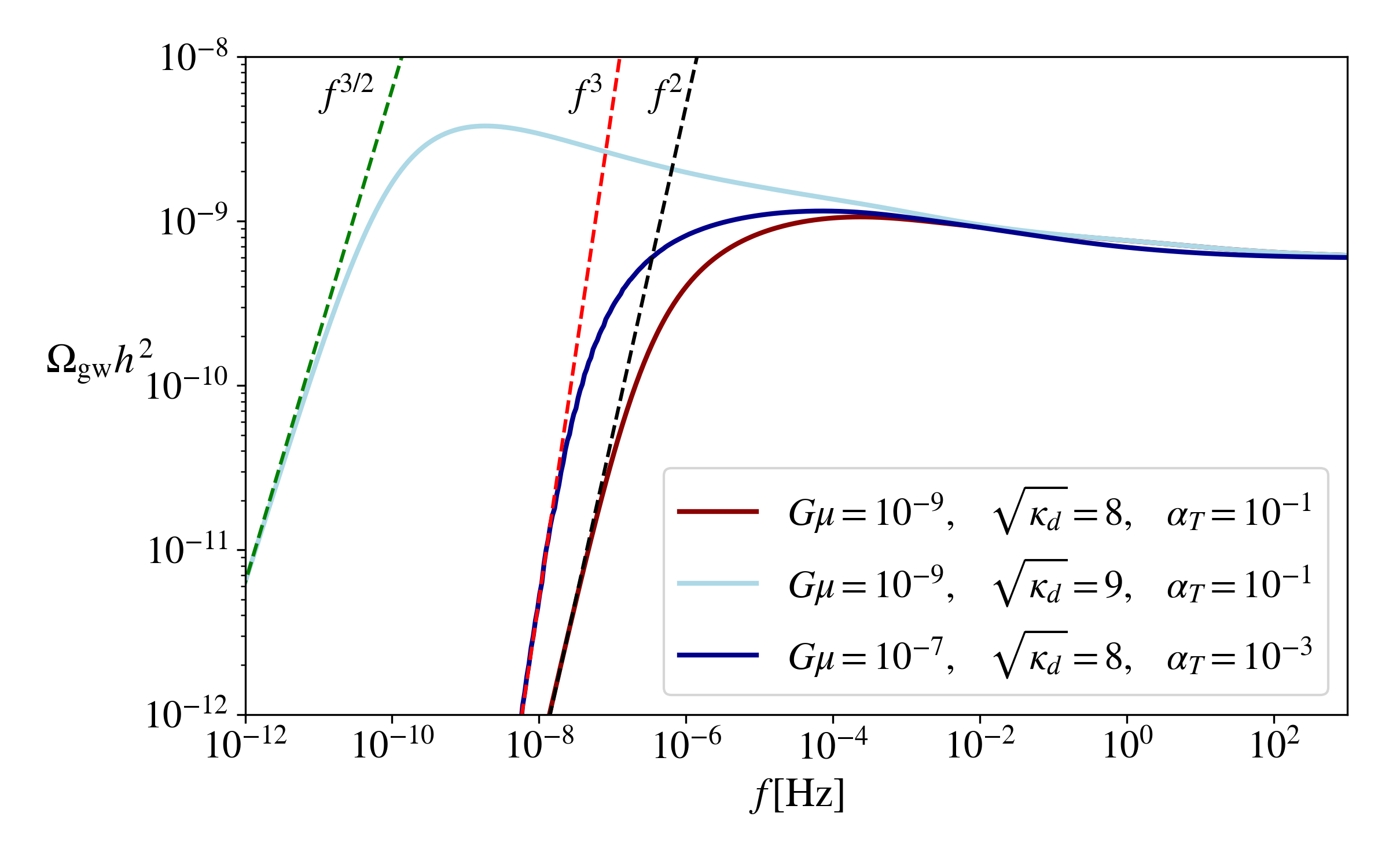}
\caption{\label{Fig.StdSlopes}
SGWB produced by MSCSs without current, illustrating the impact of varying the model parameters. In particular, the figure shows the effect of varying the loop size, string tension, and decay parameter $\kappa_d$, demonstrating how these parameters modify the low-frequency spectral slope of the SGWB, with spectral index $q=4/3$, $\mathcal{F}=0.1$ and $\alpha_T \equiv \zeta \alpha$}.
\end{center}
\end{figure}

The SGWB spectra produced by MSCSs without current, for different choices of model parameters, are shown in Fig.~\ref{Fig.StdSlopes}. The figure demonstrates that variations in cosmic string network parameters can lead to different spectral slopes at low frequencies. Recent work \cite{Asl:2026zpj} showed that an explicit separation between the timescale of the collapse of the string network and the timescale of cosmic string breaking can generate a new spectral behavior with a slope $\propto f^3$. However, as illustrated in Fig.~\ref{Fig.StdSlopes}, variations in string parameters can also modify the low-frequency slope. Furthermore, we will demonstrate below that the $f^3$ behavior is the generic asymptotic behavior of the low-frequency spectrum, while the $f^2$ behavior corresponds to a transient regime arising for particular choices of cosmic string parameters.

\subsection{SGWB for metastable current-carrying strings and vortons}
\label{III.2}

The presence of a current on cosmic strings introduces additional effects that modify the predicted SGWB. In particular, it can lead to the formation of stable current-carrying loops known as vortons \cite{Davis:1988ij}. The formation and stability of vortons have been investigated using a variety of approaches, including field-theory simulations \cite{Lemperiere:2003yt, Radu:2008pp, Battye:2008mm, Babeanu:2011ie, Garaud:2013iba, Battye:2021kbd, Battye:2021sji} and the zero-thickness approximation \cite{Carter:1993wu, Martins:1998gb, Cordero-Cid:2002hmv}, while their cosmological implications were discussed in \cite{Carter:1999an, Brandenberger:1996zp, Auclair:2020wse}.

We incorporate the effect of vorton formation through a simple phenomenological prescription. Consider a loop formed with an average charge amplitude $Y_b$, and assume that its charge is conserved throughout its subsequent evolution. The loop can then evolve toward a stationary configuration with a characteristic size \cite{Rybak:2024our} $l_{\rm vort}=l_bY_b$.
Equivalently, this condition can be expressed as conservation of the product $lY=\mathrm{const}$.

\begin{figure}[t!]
\begin{center}
\includegraphics[scale=0.65]{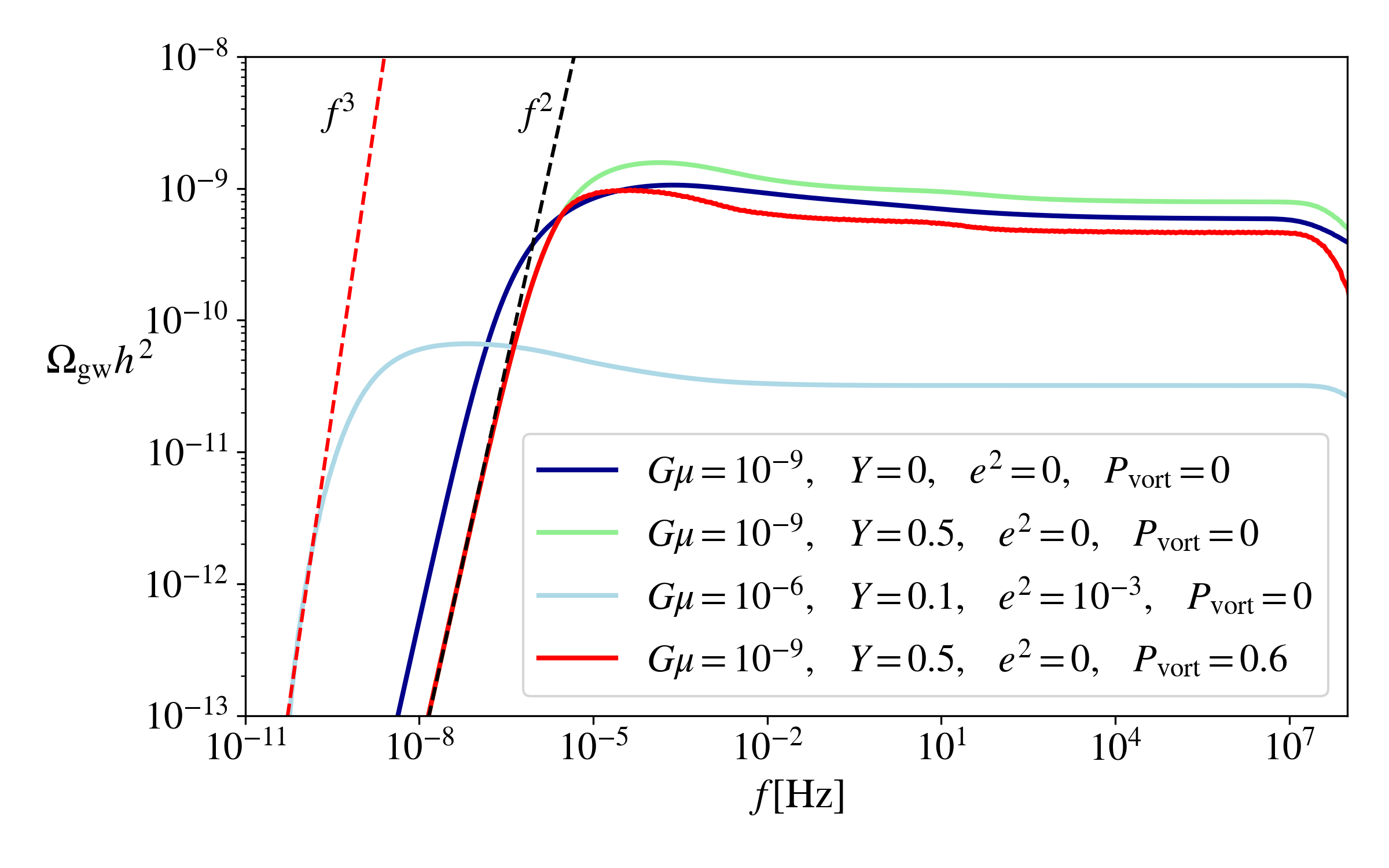}
\caption{\label{Fig.SGWB_supercond}
SGWB spectra from metastable current-carrying cosmic strings for $\sqrt{\kappa_d}=8$, $\mathcal{F}=0.1$, and $\alpha_T=0.1$, where $\alpha_T \equiv \zeta \alpha$ is a commonly used parameterization \cite{Blanco-Pillado:2013qja}. The spectra are shown for different values of the string tension, charge amplitude $Y$, vector coupling $e^2$, and $P_{\rm vort}$. The dashed lines indicate the power laws $f^3$ and $f^2$. Compared with the standard MSCSs, current-carrying strings exhibit both a shift in the characteristic frequency and a change in the spectral amplitude. For the parameters considered here, vorton formation predominantly affects the high-frequency part of the spectrum, while leaving the low-frequency behavior nearly unchanged. Following \cite{Rybak:2024our}, we take the spectral index to be $q=4/3$ for $Y=0$ and $q=5/3$ for $Y\neq 0$.}
\end{center}
\end{figure}

In realistic particle physics models, however, the current carried by a string can decrease through several mechanisms \cite{Barr:1987ij,Blanco-Pillado:2002vwq,Lemperiere:2002en,Ibe:2021ctf,Abe:2022rrh}. Hence, whether a given loop ultimately reaches a stable vorton configuration depends on the microscopic dynamics of the current and on the stability properties of the resulting configuration.

A newly formed current-carrying loop is generally expected to be non-circular and irregular, with sharp features such as kinks. These features can enhance processes that lead to charge loss \cite{Ibe:2021ctf}, which can affect the charge carried by the loop as it evolves toward a stationary configuration. During its subsequent evolution, gravitational and current-related backreaction can smooth the loop and drive it toward a more regular configuration, which eventually becomes a vorton. A first-principles description of this process is highly nonlinear. Although simulations confirm the possibility of vorton formation and stability \cite{Garaud:2013iba, Battye:2021kbd, Battye:2021sji}, the dynamical formation of vortons from a string network remains an open question.

We incorporate the possibility of vorton formation, as studied in Ref.~\cite{Auclair:2020wse}. To describe the uncertainty associated with the dynamical formation process, we introduce a parameter $0 \leq P_{\rm vort}\leq1$ that characterizes the fraction of the characteristic loop size $l_bY_b$ that remains when the loop reaches a stable vorton configuration. We therefore define the characteristic size of the resulting vorton as
\begin{equation}
l_{\rm vort} = P_{\rm vort}  l_b  Y_b
= P_{\rm vort}  Y_b  \alpha  \zeta_b  t_b .
\end{equation}
Here, $P_{\rm vort}=1$ corresponds to the idealized case in which the loop reaches the vorton configuration without additional length loss, and the characteristic size is $l_bY_b$. Smaller values of $P_{\rm vort}$ correspond to a smaller vorton size. By varying $P_{\rm vort}$, we phenomenologically account for the uncertainty in the dynamical formation of vortons and study its impact on the predicted SGWB.

As the loop shrinks, the current amplitude $Y$ is expected to increase and eventually reach its maximum value, $Y=1$, at the moment when the vorton is formed. A consistent treatment requires following the continuous evolution of $Y$ throughout this process. Instead, we adopt a simplified step-function approximation in which $Y$ is kept fixed at its initial value $Y_b$ during the loop evolution and is set to $Y=1$ when the loop reaches its final vorton size. This approximation introduces some uncertainty in the detailed dynamics of vorton formation. However, we find that accounting for the continuous evolution of $Y$ does not lead to significant changes in the predicted SGWB, while requiring much more computationally expensive calculations.

Figure~\ref{Fig.SGWB_supercond} demonstrates the effects of the current and vorton formation on the SGWB. Compared with the standard NG case, the presence of a current modifies both the characteristic frequency and the amplitude of the spectrum. An enhanced total decay rate $\Gamma_{\rm tot}$ due to vector coupling suppresses the overall amplitude of the SGWB signal and can lead to a steeper low-frequency spectrum, approaching an $f^3$ behavior earlier than in the NG case.

Vorton formation has a qualitatively different effect on the SGWB. Once a loop becomes a vorton, it stops radiating, which reduces the high-frequency contribution since loops cannot radiate after reaching the characteristic size of a vorton. This effective cutoff is parametrized by $P_{\rm vort}$ in our study. As illustrated in Fig.~\ref{Fig.SGWB_supercond}, a non-trivial value of $P_{\rm vort}$ changes the high-frequency amplitude while leaving the low-frequency spectrum nearly unchanged.

\section{Analytical template for SGWB}
\label{IV}

We aim to develop a computationally efficient template for the SGWB generated by current-carrying MSCS networks, using a phenomenological approach that is agnostic to the specific parameters of the network evolution. This allows us to consider more general and realistic scenarios with additional degrees of freedom for which string network simulations are not yet available to determine the loop number density and, consequently, the resulting SGWB. Our goal is to construct a phenomenological description of the SGWB in terms of free parameters that can capture possible modifications to the evolution of cosmic string networks beyond the NG approximation. In particular, we investigate how such modifications affect the agreement with current PTA data.

We focus on current-carrying MSCS networks that decay before the radiation-matter equality epoch. This scenario is of particular interest because it can provide a viable contribution to the PTA signal while remaining consistent with CMB constraints.

During radiation domination, when $a \propto t^{1/2}$, we assume that the string network is in the scaling regime, such that $Y$ and $v$ are constant and $\xi_r = t \zeta$. Under these assumptions, the time of loop formation can be obtained from \eqref{LengthChange} as
\begin{equation}
t_b = \frac{l + \Gamma_{\rm tot} t}{\zeta \alpha + \Gamma_{\rm tot}}.
\end{equation}
Substituting this expression for $t_b$ into \eqref{LoopNumbDensG2}, and using the corresponding expressions for the scaling network parameters, we obtain
\begin{equation}
\label{LoopNumberDensFin}
n(l,t) = \frac{\mathcal{F} C_r(\alpha) }{t^{3/2} \left( \ell + t \right)^{5/2}} \text{e}^{- \Gamma_d \Gamma_{\rm tot} \left[ \ell \left( t- \frac{\ell + t}{\epsilon_r + 1} \right) + \frac{1}{2} \left(t- \frac{\ell + t}{\epsilon_r + 1} \right)^2 \right] },
\end{equation}
where we have introduced the notation
\begin{equation}
C_r(\alpha) \equiv c v \sqrt{1-v^2} \frac{\left( \epsilon_r + 1 \right)^{3/2}}{ \Gamma_{\rm tot} \alpha \zeta^4},
\end{equation}
and defined
\begin{equation}
\ell \equiv \frac{l}{\Gamma_{\rm tot}}, \qquad
\epsilon_r \equiv \frac{\alpha \zeta}{\Gamma_{\rm tot}},
\end{equation}
with $l = \frac{2 j}{f} \frac{a(t)}{a_0} S(Y)$.

In the limit $\epsilon_r \gg 1$, the expression \eqref{LoopNumberDensFin} reduces to the result obtained in \cite{Buchmuller:2021mbb}. This condition are generally satisfied for standard NG strings when the contribution from small loops can be neglected \cite{Blanco-Pillado:2013qja, Blanco-Pillado:2019tbi}. However, this approximation may no longer be valid when considering deviations from the NG description and allowing a parameter range in which $\epsilon_r \sim \mathcal{O}(1)$. In particular, for superconducting cosmic strings \cite{Rybak:2024our}, $\Gamma_{\rm tot}$ can be significantly larger due to the additional vector radiation channel, which can have a much larger coupling constant $e^2$. The enhanced energy-loss rate into this additional channel can increase $\Gamma_{\rm tot}$ and reduce $\epsilon_r = \alpha \zeta/\Gamma_{\rm tot}$ to values of order unity, making the $\epsilon_r \gg 1$ approximation no longer applicable. For the case of electromagnetic radiation from superconducting strings, additional constraints can be imposed \cite{Cyr:2023iwu, Afzal:2026ost}.

\subsection{Fundamental modes}
\label{4.1}

In this subsection, we derive an analytical approximation for the contribution of the fundamental mode to the SGWB produced by current-carrying MSCSs. We follow the approach of Ref.~\cite{Sousa:2020sxs}, where analogous analytical expressions were derived for standard NG cosmic strings. According to Eq.~\eqref{t_s}, the cosmic string network can be treated as effectively stable at times $t \lesssim t_s$, where $t_s \approx \Gamma_d^{-1/2}$.
In this regime, the exponential suppression due to string decay in the loop number density \eqref{LoopNumberDensFin} can be neglected. We therefore first consider the contribution from the period when the decay exponent is negligible and subsequently determine the appropriate upper cutoff for the integration.

For a scaling string network during radiation domination, we can then use
\begin{equation}
\label{Simple}
n_r(l,t) =  \frac{\mathcal{F} C_r (\alpha) }{t^{3/2} \left( \ell + t \right)^{5/2}}.
\end{equation}
Carrying out the integration in Eq.~\eqref{Omega_j} explicitly for a scaling network during radiation domination, and setting $a_0=1$ and $j=1$, we obtain
\begin{equation}
    \label{OmegaR}
    \Omega^{r0}_{\rm gw} (f) = \frac{128 \pi}{9} \frac{ \mathcal{F} C_r (\alpha) }{S(Y)} \frac{(G \mu)^2 \Gamma(Y)} \Gamma_{\rm tot} {\mathcal{E}(Y)  } \Omega_r \left( \frac{ 1 }{1+B_r/a} \right)^{3/2} \Bigg|_{a=a_{\rm min}}^{a=a_{\rm fr}},
\end{equation}
where
\begin{equation}
B_r = \frac{4 H_0 S(Y) \Omega_r^{1/2}}{\Gamma_{\rm tot} f} \qquad \text{and} \qquad a_{\rm min} = \frac{ \left( 2 H_0 \Omega_r^{1/2} \left( \epsilon_r + 1 \right) t_{\rm pl} \right)^{1/2} }{G \mu},
\end{equation}
Here, $a_{\rm fr}$ denotes the upper limit of the integration, which is determined by the finite lifetime of the MSCS network and/or by the exponential suppression of the loop population and $t_{\rm pl}$ is the Planck time. We consider the possible mechanisms determining this cutoff separately below.

\subsubsection{Exponential cutoff}
\label{ExponentCutOff}

First, let us assume that the dominant contribution to the exponent in Eq.~\eqref{LoopNumberDensFin} comes from its first term. Following the argument of \cite{Buchmuller:2021mbb}, we can then estimate the scale factor at which the exponential suppression becomes important as
\begin{equation}
\label{CutOff1}
a_{\rm fr} = \left( \frac{ H_0 \Omega_r^{1/2} (1+\epsilon_r) f }{\Gamma_d \epsilon_r S(Y) j } \right)^{1/3}.
\end{equation}
Substituting Eq.~\eqref{CutOff1} into Eq.~\eqref{OmegaR} and considering the low-frequency regime, for which $B_r/a_{\rm fr} \gg 1$, we find
\begin{equation}
\Omega^r_{\rm gw}(f) \sim f^2.
\end{equation}
This cutoff reproduces the familiar $f^2$ behavior obtained for standard NG cosmic strings in \cite{Buchmuller:2021mbb}.

\subsubsection{Minimum frequency cutoff}
\label{MinimFreq}

A second characteristic frequency arises from the finite lifetime of the string network. The largest loops, and therefore those emitting at the lowest frequencies, are produced at the latest times, when the network is still present. We therefore consider a loop produced at $t=t_s$, with initial length $\alpha\zeta t_s$. The corresponding minimum frequency is determined by
\begin{equation}
2 \frac{a_s j S(Y)}{a_0 f_{\rm min}} = \alpha \zeta t_s,
\end{equation}
where $a_s$ is the scale factor at $t=t_s$. Using $t_s = 1/\sqrt{\Gamma_d}$ and taking the fundamental mode, $j=1$, which gives the smallest frequency, we obtain
\begin{equation}
\label{FreqMin}
f_{\rm min} = \frac{2 S(Y) \sqrt{ 2 H_0 \Omega_r^{1/2} } \Gamma_d^{1/4}}{ \alpha \zeta }.
\end{equation}
Eq.~\eqref{FreqMin} means that there is no contribution to the SGWB from the string network at frequencies below $f_{\rm min}$.

\subsubsection{String network decay cutoff}
\label{StrNetworkDecay}

The upper cutoff $a_{\rm fr}$ is determined not only by the exponential suppression of the loop number density, but also by the finite lifetime of the string network. A similar distinction between the timescale of network decay and the loop-breaking timescale was introduced phenomenologically for NG strings in Ref.~\cite{Asl:2026zpj}. In our case, the corresponding cutoff can be estimated directly from the condition
\begin{equation}
l = 2 \frac{a j S(Y)}{a_0 f} = \alpha \zeta t_s - \Gamma_{\rm tot}(t-t_s).
\end{equation}
This condition determines the maximum scale factor up to which loops produced near the time of string network decay can still contribute to the SGWB at a given frequency and mode. Substituting $t_s = 1/\sqrt{\Gamma_d}$ and the radiation-era relation between $t$ and $a$, we obtain
\begin{equation}
\label{CutOff2}
a_{\rm fr} = H_0\sqrt{\Omega_r}
\left(
 \sqrt{
\frac{2 (\epsilon_r+1)}
{H_0\sqrt{\Omega_r \Gamma_d}}
+\frac{4 j^2 S(Y)^2}{f^2 \Gamma_{\rm tot}^2}
}
-\frac{2 j S(Y)}{f \Gamma_{\rm tot}}
\right).
\end{equation}

At sufficiently low frequencies and/or for sufficiently high modes, the second term inside the square root dominates, such that
\begin{equation}
\frac{2 (\epsilon_r+1)}
{H_0\sqrt{\Omega_r \Gamma_d}}
\ll
\frac{4 j^2 S(Y)^2}{f^2 \Gamma_{\rm tot}^2}.
\end{equation}
In this limit, Eq.~\eqref{CutOff2} reduces to
\begin{equation}
\label{CutOff2Asympt}
a_{\rm fr} =  \frac{(1+\epsilon_r) f \Gamma_{\rm tot} }
{2 \sqrt{\Gamma_d} S(Y) j},
\end{equation}
which has a form similar to the result obtained in Eq.~(34) of Ref.~\cite{Asl:2026zpj}.

The transition between the two cutoff regimes, given by Eqs.~\eqref{CutOff1} and \eqref{CutOff2Asympt}, can be characterized by a frequency $f_j^*$ defined by equating the two contributions. For each mode $j$, this gives
\begin{equation}
\label{FrChar}
f_j^* \equiv \frac{f^*}{j} \approx
\frac{2 \sqrt{2} S(Y)}
{\Gamma_{\rm tot} + \alpha \zeta}
\sqrt{
\frac{H_0 \sqrt{\Omega_r \Gamma_d}}
{\alpha \zeta}
}.
\end{equation}
We can therefore use
\begin{equation}
    \label{CuttOffF}
    a_{\rm fr} = 
    \begin{cases}
         \eqref{CutOff2}, \quad \text{when} \quad f/j<f^*_j \\
         \eqref{CutOff1}, \quad \text{when} \quad f/j>f^*_j.
    \end{cases}
\end{equation}
In practice, the relevant upper cutoff is determined by whichever mechanism suppresses the loop contribution first.

Substituting Eq.~\eqref{CutOff2Asympt} into Eq.~\eqref{OmegaR} and again considering the asymptotic regime $B_r/a_{\rm fr} \gg 1$, we obtain
\begin{equation}
\label{f^3}
\Omega^r_{\rm gw}(f) \sim f^3.
\end{equation}
The two different cutoff mechanisms give rise to two distinct power-law behaviors: the exponential cutoff yields the $f^2$ regime, whereas the string network decay cutoff leads to the $f^3$ regime.

\begin{figure}[t!]
\begin{center}
\includegraphics[scale=0.65]{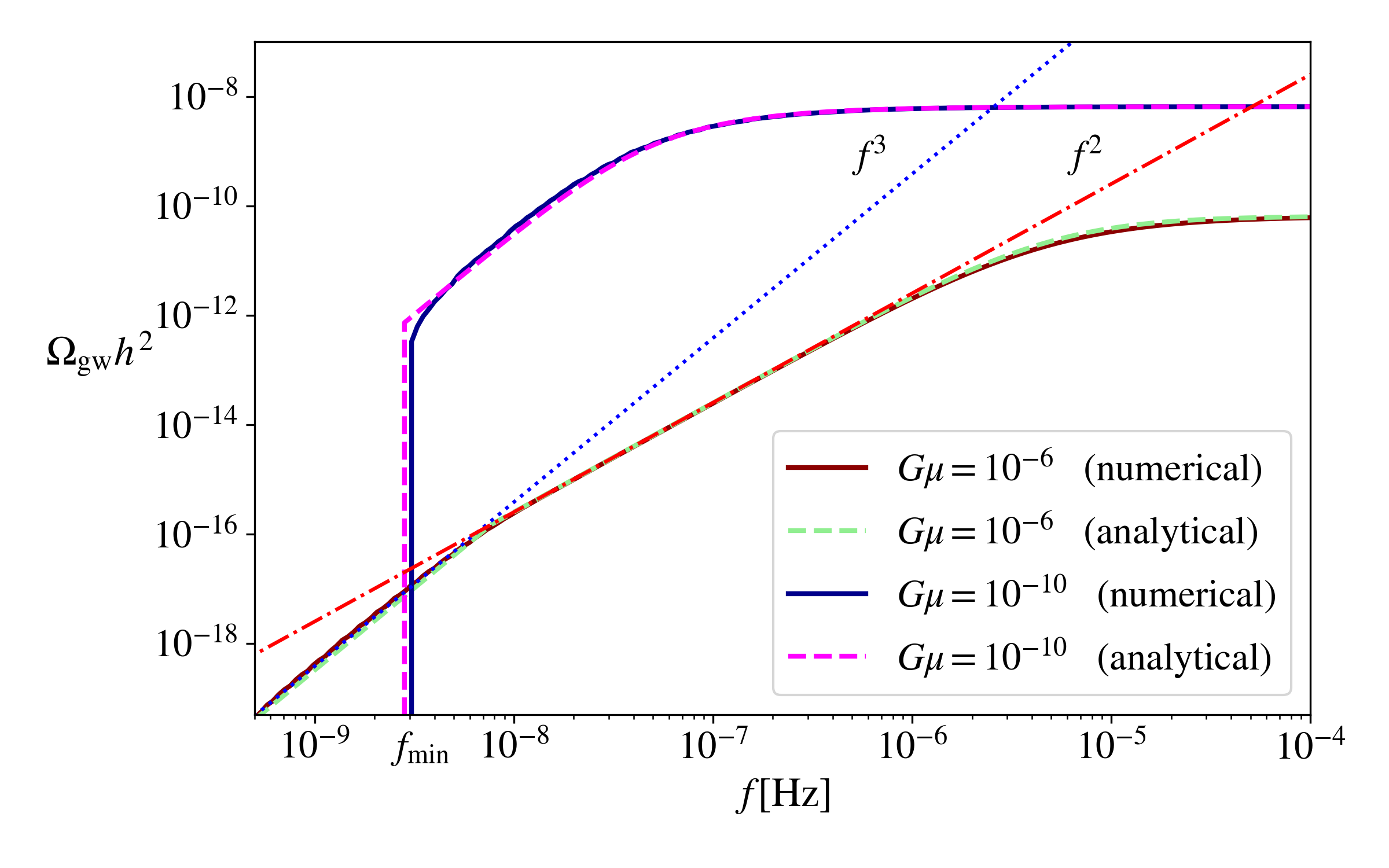}
\caption{\label{Fig.SGWB_comparisonNA}
Comparison between the analytical approximation and the full numerical calculation of the SGWB produced by metastable current-carrying cosmic strings. The fundamental mode is shown for $Y=0.5$, $\sqrt{\kappa_d}=8$, $\alpha_T=10^{-3}$, $\mathcal{F}=0.1$, and $e^2=0$, for two different values of $G\mu$. The two characteristic regimes with slopes $f^2$ and $f^3$ are visible, with the transition between them determined by Eq.~\eqref{FrChar}.
}
\end{center}
\end{figure}

This result also shows that an explicit separation between the timescale of string network decay and the loop-breaking timescale, as introduced phenomenologically in \cite{Asl:2026zpj}, is not required to obtain the asymptotic $f^3$ behavior. In particular, if the minimum frequency $f_{\rm min}$ given by Eq.~\eqref{FreqMin} is smaller than the transition frequency $f_1^*$ of the fundamental mode, the $f^3$ regime is always present. The analytical approximation for the fundamental mode and its agreement with the full numerical calculation are demonstrated in Fig.~\ref{Fig.SGWB_comparisonNA}, which also illustrates the transition from the $f^2$ to the $f^3$ regime for a particular choice of parameters.

\subsection{Parameter dependence of the SGWB}

The analytical expression \eqref{OmegaR} allows us to identify the main effects of the cosmic string network parameters on the SGWB. Increasing the gravitational radiation efficiency $\Gamma$ generally suppresses the SGWB amplitude. A larger $\Gamma$ causes loops to lose their energy more rapidly, shifting a larger fraction of gravitational wave emission to earlier times, when the resulting radiation is more strongly suppressed by cosmological redshift \cite{Rybak:2022sbo}. Similarly, increasing the efficiency of vector radiation suppresses the SGWB by increasing the fraction of energy lost through the vector radiation channel rather than through gravitational waves \cite{Rybak:2024our}.

The dependence on the loop-size parameter $\alpha$ is more subtle. As $\alpha$ decreases, the SGWB amplitude initially decreases until $\alpha$ approaches the critical value
\begin{equation}
\alpha_{\rm cr} = \frac{2\Gamma_{\rm tot}}{\zeta}.
\end{equation}
For $\alpha < \alpha_{\rm cr}$, the analytical expression predicts an increase of the SGWB amplitude, which eventually diverges as $\alpha \rightarrow 0$. This behavior signals the breakdown of the underlying long-string evolution model: in this limit, energy loss through radiation becomes increasingly important relative to loop production, while the corresponding gravitational and vector backreaction on the long-string network is not included in the evolution equations \eqref{EqOfMotMacroLin2}. Therefore, the region $\alpha \lesssim \alpha_{\rm cr}$ should be interpreted with caution.

Finally, the network parameters also affect the characteristic slope of the spectrum at low frequencies. In particular, the spectrum at low frequencies can exhibit a transient $\Omega^r_{\rm gw} \sim f^2$ regime, followed by the asymptotic $\Omega^r_{\rm gw} \sim f^3$ behavior at even lower frequencies, as discussed above. The corresponding transition frequencies depend on the network parameters through the cutoff scales derived in subsection \ref{4.1}.

\subsubsection{Relativistic degrees of freedom}

To account for the effect of the change in the number of relativistic degrees of freedom (DOF) on the evolution of the string network and the resulting SGWB \cite{Caldwell:1996en}, we approximate the evolution of the effective number of DOF by three step functions. This allows us to evaluate the SGWB integral piecewise over the corresponding intervals in the scale factor: 
\begin{equation}
\begin{gathered}
    \label{OmegaR0DOF}
    \Omega^{r 0}_{\rm gw} (f) = \mathcal{C}_r  \Bigg[   \left( \frac{ 1 }{1+B_r /a_{\rm fr}} \right)^{3/2} - \left( \frac{ 1 }{1+B_r /a_{3}} \right)^{3/2}   \\
    + \Delta g_3^* \left( \left( \frac{ 1 }{1+ (\Delta g_3^*)^{1/2} B_r /a_{3}} \right)^{3/2} - \left( \frac{ 1 }{1+(\Delta g_3^*)^{1/2} B_r /a_{2}} \right)^{3/2}  \right)  \\
    + \Delta g_2^* \left( \left( \frac{ 1 }{1+ (\Delta g_2^*)^{1/2} B_r /a_{2}} \right)^{3/2} - \left( \frac{ 1 }{1+(\Delta g_2^*)^{1/2} B_r /a_{1}} \right)^{3/2}  \right) \\
    + \Delta g_1^* \left( \left( \frac{ 1 }{1+ (\Delta g_1^*)^{1/2} B_r /a_{1}} \right)^{3/2} - \left( \frac{ 1 }{1+ (\Delta g_1^*)^{1/4} B_r /a_{\rm min}} \right)^{3/2} \right) \Bigg] ,
\end{gathered}
\end{equation}
following the approach of \cite{Blanco-Pillado:2024aca}, where $\mathcal{C}_r =  \frac{128 \pi}{9} \frac{ \mathcal{F} C_r (\alpha) }{S(Y)} \frac{(G \mu)^2 \Gamma \Gamma_{\rm tot} }{\mathcal{E}(Y)  } \Omega_r$. The precise form of the resulting expression depends on the position of the string network decay cutoff $a_{\rm fr}$ relative to the transition scale factors $a_1$, $a_2$, and $a_3$.

Here, $a_1$, $a_2$, and $a_3$ denote the scale factors at which the effective number of relativistic DOF changes, while $\Delta g_i^*$ parameterizes the corresponding change in the number of DOF at each transition. The first term in Eq.~\eqref{OmegaR0DOF} corresponds to the interval $[a_3,a_{\rm fr}]$, while the remaining terms account for the contributions from the other three intervals. The same procedure can be applied to any location of the decay cutoff, $a_{\rm fr}<a_3$, by terminating the piecewise integration at the corresponding value of $a_{\rm fr}$ with the corresponding factor $\Delta g_i^*$.

As in \cite{Blanco-Pillado:2024aca}, the parameters $a_i$ and $\Delta g_i^*$ are fitted to the results of the full numerical integration described in Section~\ref{III}. We obtain $\Delta g_1^* = 0.397$, $\Delta g_2^* = 0.488$,
$\Delta g_3^* = 0.724$ and $a_1 = 3.717\times10^{17} G\mu \, a_{\rm min}$, $a_2 = 6.951 \times 10^{19} G\mu \, a_{\rm min}$, $a_3 = 1.364 \times 10^{23} G\mu \, a_{\rm min}$.

\subsubsection{Effect of vortons}

To incorporate the effect of vorton formation discussed in Section~\ref{III.2} into the analytical treatment, we derive an additional cutoff on the loop contribution. From Eq.~\eqref{LengthChange}, the condition that a loop remains larger than the corresponding vorton size, $l \geq l_{\rm vort}$, gives
\begin{equation}
\label{ScaleFactVort}
   \frac{ 4 H_0 \Omega_r^{1/2} S(Y) \left( \epsilon_r \left( 1 - P_{\rm vort} \, Y_b \right) + 1 \right) }{P_{\rm vort} \, Y_b \, \epsilon_r \, \Gamma_{\rm tot} }  \frac{j}{f} \equiv \frac{a_{\rm vort}}{a_0} \ge \frac{a }{a_0}.
\end{equation}
Here, $a_{\rm vort}$ denotes the maximum scale factor up to which loops of a given frequency and mode can contribute before their length falls below the corresponding vorton size. Therefore, the upper limit of the SGWB integration must be modified whenever the vorton cutoff occurs before the cutoff associated with MSCS decay, namely when $a_{\rm vort} < a_{\rm fr}$. In this case, the effective upper integration limit is replaced by $a_{\rm vort}$. When the change in the relativistic DOF is taken into account, the corresponding $a_{\rm vort}$ is modified by the appropriate factor $\Delta g_i^*$ depending on the interval in which the upper integration limit lies.

The vorton cutoff is expected to be particularly relevant for the high-frequency contribution, for which the corresponding loops are typically smaller. In the limit $P_{\rm vort}\rightarrow 0$, the vorton constraint becomes ineffective and the standard scenario is recovered.

\subsection{Complete computational template for the SGWB}

To obtain the full SGWB contribution from all modes of cosmic string loops, we need to perform the summation over $j$ up to a maximum mode number $N$, which accounts for the truncation due to gravitational backreaction and the finite string thickness, as specified in Eq.~\eqref{OmegaSGWB}. Since $N$ can be very large, we fix it to $N=10^5$ \cite{Sanidas:2012ee}. Explicitly summing all modes up to this value can significantly increase the computational cost.

\begin{figure}[t!]
\begin{center}
\includegraphics[scale=0.65]{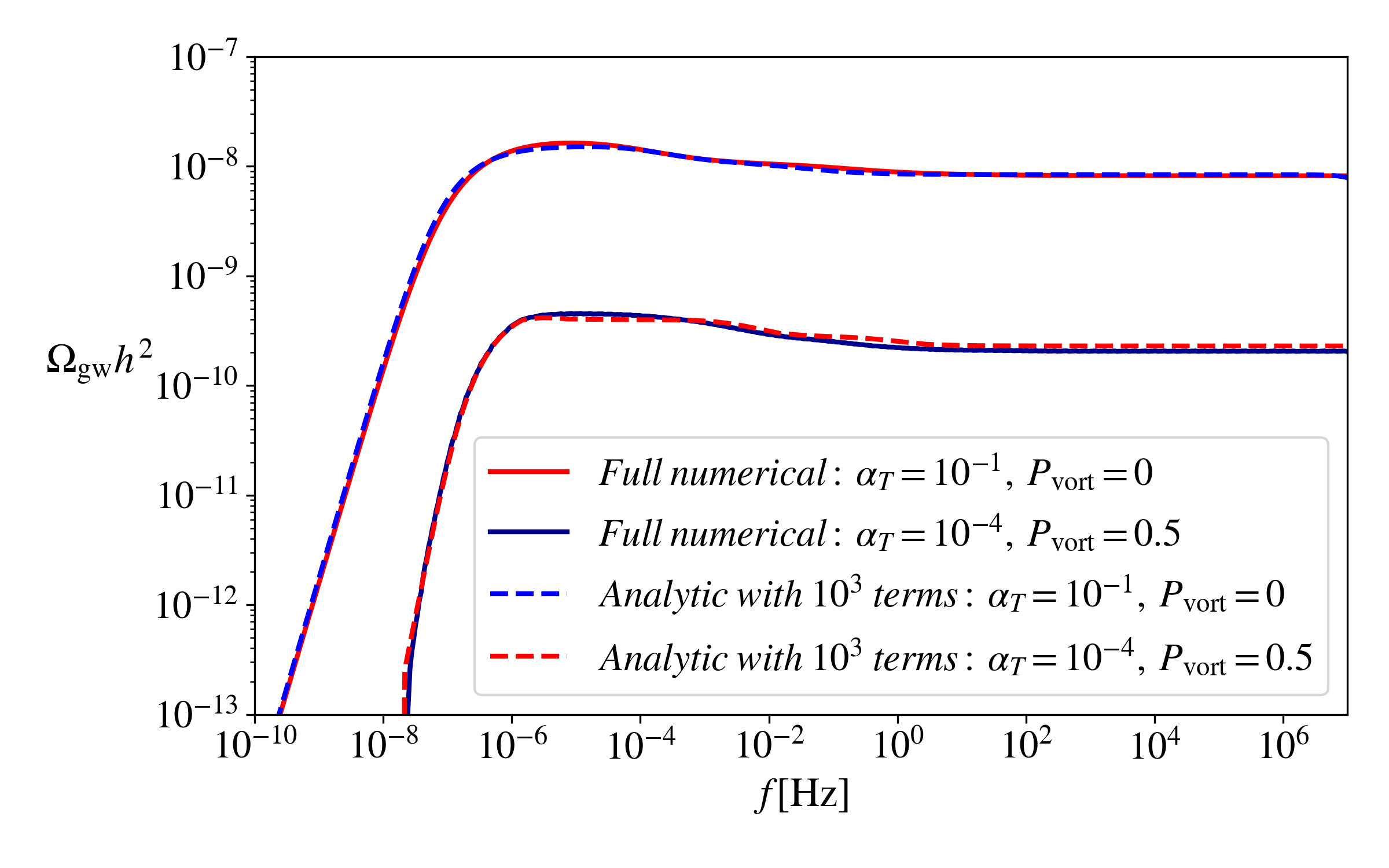}
\includegraphics[scale=0.65]{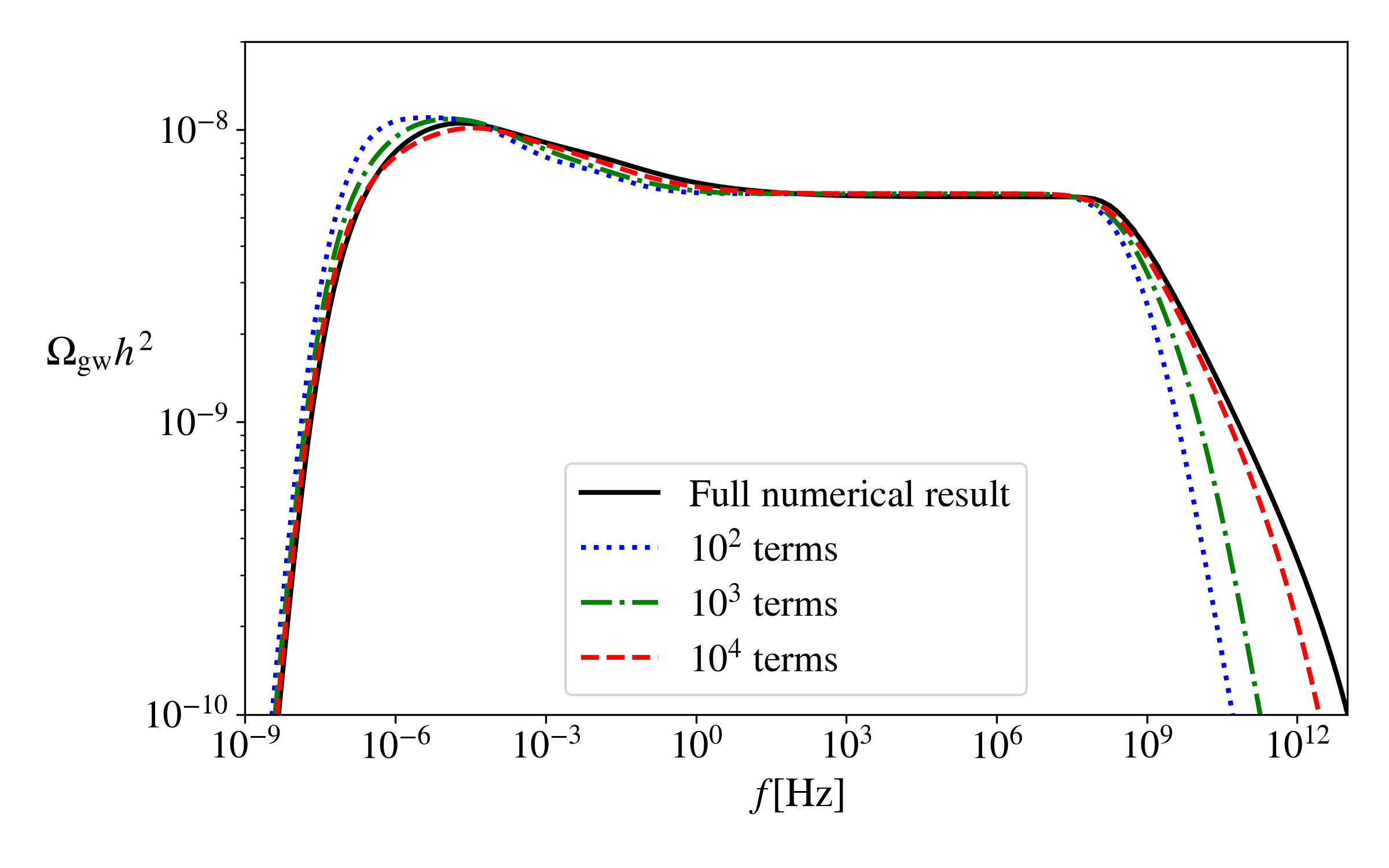}
\caption{\label{Fig.SGWB_comparisonNA2}
\textit{Upper panel:} Comparison between the analytical approximation and the full numerical calculation of the SGWB produced by current-carrying MSCSs. The numerical calculation includes $10^5$ terms, while the analytical approximation includes $10^3$ terms, using the following cosmic string parameters: $Y=0.3$, $\sqrt{\kappa_d}=8$, $G \mu = 10^{-7}$, $\mathcal{F}=0.1$, and $e^2=0$. The approximation reproduces the characteristic $f^2$ and $f^3$ slopes, together with the changes in the relativistic degrees of freedom that create small bumps. \\
\textit{Lower panel:} Comparison between the full numerical spectrum and the analytical approximation with different numbers of terms in the summation for standard cosmic strings with cusps, corresponding to $q=4/3$, $G\mu=10^{-7}$, $\sqrt{\kappa_d}=8$, $\mathcal{F}=0.1$ and $\alpha_T = 0.1$. As the number of terms increases, the analytical approximation converges toward the full numerical result. The convergence is slower for cusps ($q=4/3$) than for kinks ($q=5/3$), requiring more terms to achieve the same level of accuracy.
}
\end{center}
\end{figure}

For quasi-cusps, the exponential factor in Eq.~\eqref{gCK} rapidly suppresses the contribution of higher modes. Therefore, the summation need not be explicitly performed up to $N$; instead, it can be truncated once the desired numerical precision is reached. Kinks, however, provide the dominant contribution for current-carrying strings and have a spectrum that decreases more slowly with increasing mode number. Their contribution, therefore, requires a more careful treatment of the high-mode tail.

For the present study, we are primarily interested in explaining the PTA data with the SGWB generated by MSCSs. In this frequency range, the dominant contribution comes from the lowest modes of the largest loops, which emit gravitational waves in the frequency range probed by PTAs. Therefore, when we construct the model for \texttt{PTArcade} \cite{Mitridate:2023oar}, we can restrict the number of explicitly summed modes while maintaining high accuracy. In particular, we sum explicitly over the first $10^3$ modes for kinks and the first $10^4$ modes for cusps; see Fig.~\ref{Fig.SGWB_comparisonNA2} for a visual comparison.

The explicit summation over these modes can be evaluated efficiently using vectorized operations and is computationally faster than the Euler–Maclaurin approximation employed in \cite{Blanco-Pillado:2024aca} when only the low-frequency part of the spectrum is of interest. To obtain more accurate predictions at high frequencies, the contribution from the remaining high-frequency modes can be approximated using the Euler–Maclaurin formula \cite{Blanco-Pillado:2024aca}. This hybrid approach reduces the computational cost while accounting for the high-mode tail and providing an accurate approximation to the full SGWB spectrum.

\section{Observational constraints}
\label{V}

Cosmic strings are subject to a variety of observational constraints. Among the most stringent constraints on stable cosmic string networks are those obtained from their contribution to the CMB anisotropies \cite{Lazanu:2014eya, Lizarraga:2016onn, Ringeval:2015ywa, Rybak:2021scp, Raidal:2026cpb}. For standard NG cosmic strings, these analyses constrain the string tension to $G \mu < 3.66 \times 10^{-8}$ \cite{Raidal:2026cpb}. For current-carrying strings, the corresponding constraint on $G \mu$ can be weakened by approximately 25\%, for $Y \approx 0.68$. However, when the current approaches its maximal value, $Y > 0.9$, stronger constraints on $G \mu$ can be expected due to the presence of the current \cite{Rybak:2024djq}.

\begin{figure}[t!]
\begin{center}
\includegraphics[scale=0.85]{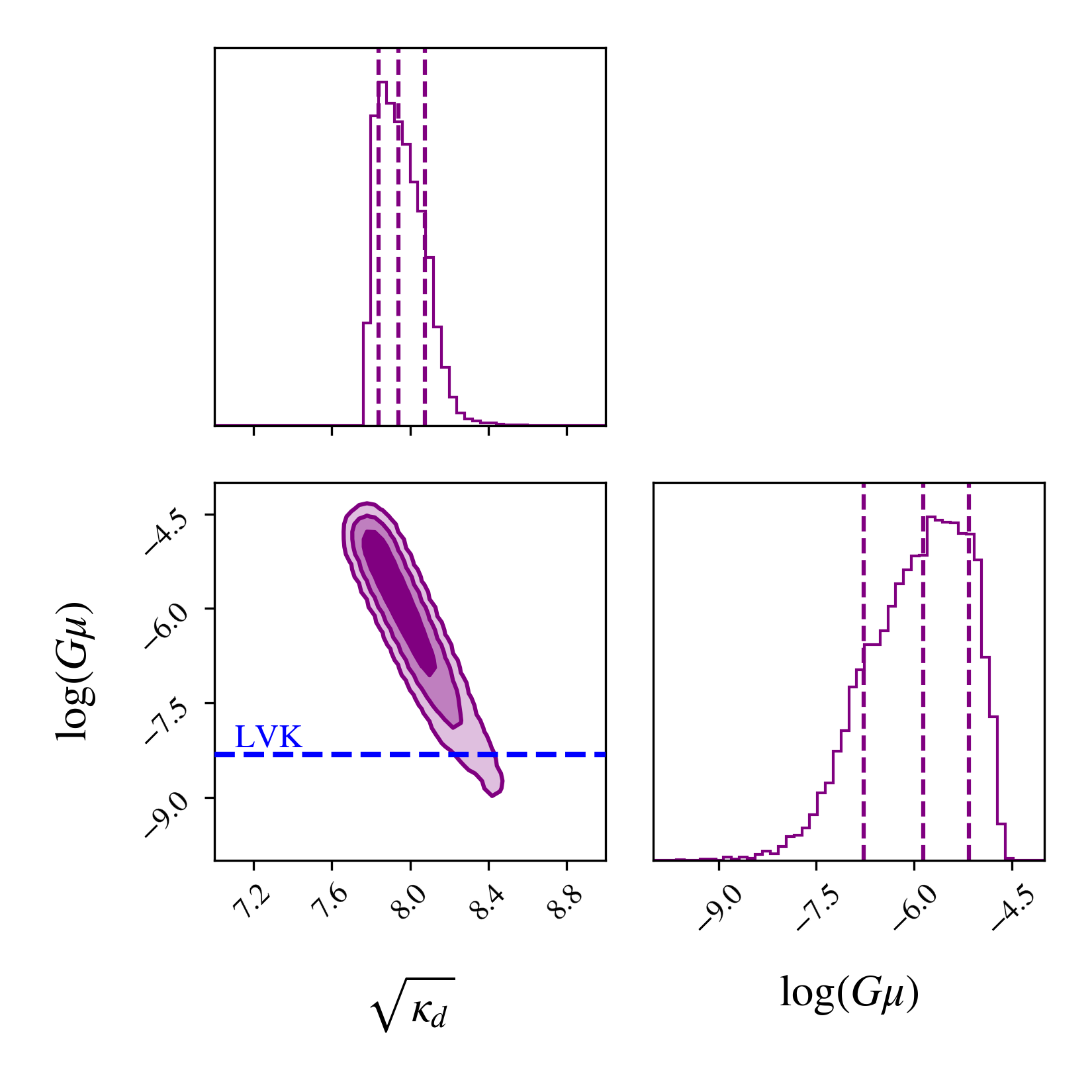}
\caption{\label{Fig.PTArcade} Results of the analysis with \texttt{PTArcade} for standard MSCSs, with uniform prior distributions $\sqrt{\kappa_d} \in \left[7,9\right]$ and $\log(G\mu) \in \left[-10,-4\right]$. The LVK constraint at 95\% credibility, $\Omega_{\rm gw}(25\,{\rm Hz}) < 2.8 \times 10^{-9}$ \cite{LIGOScientific:2025bgj}, is shown by the dashed line and is given by \eqref{LVK_std_Constr}.  }
\end{center}
\end{figure}

These constraints can be eliminated for MSCSs if the string network decays sufficiently early. In the scenario considered here, the cosmic string network decays well before the radiation-matter equality epoch and therefore long before recombination. As a result, the string network does not survive until recombination, and its direct contribution to the CMB anisotropies is suppressed. This allows us to consider regions of parameter space that would be excluded by CMB observations for stable cosmic string networks.

\begin{figure}[t!]
\begin{center}
\includegraphics[scale=0.63]{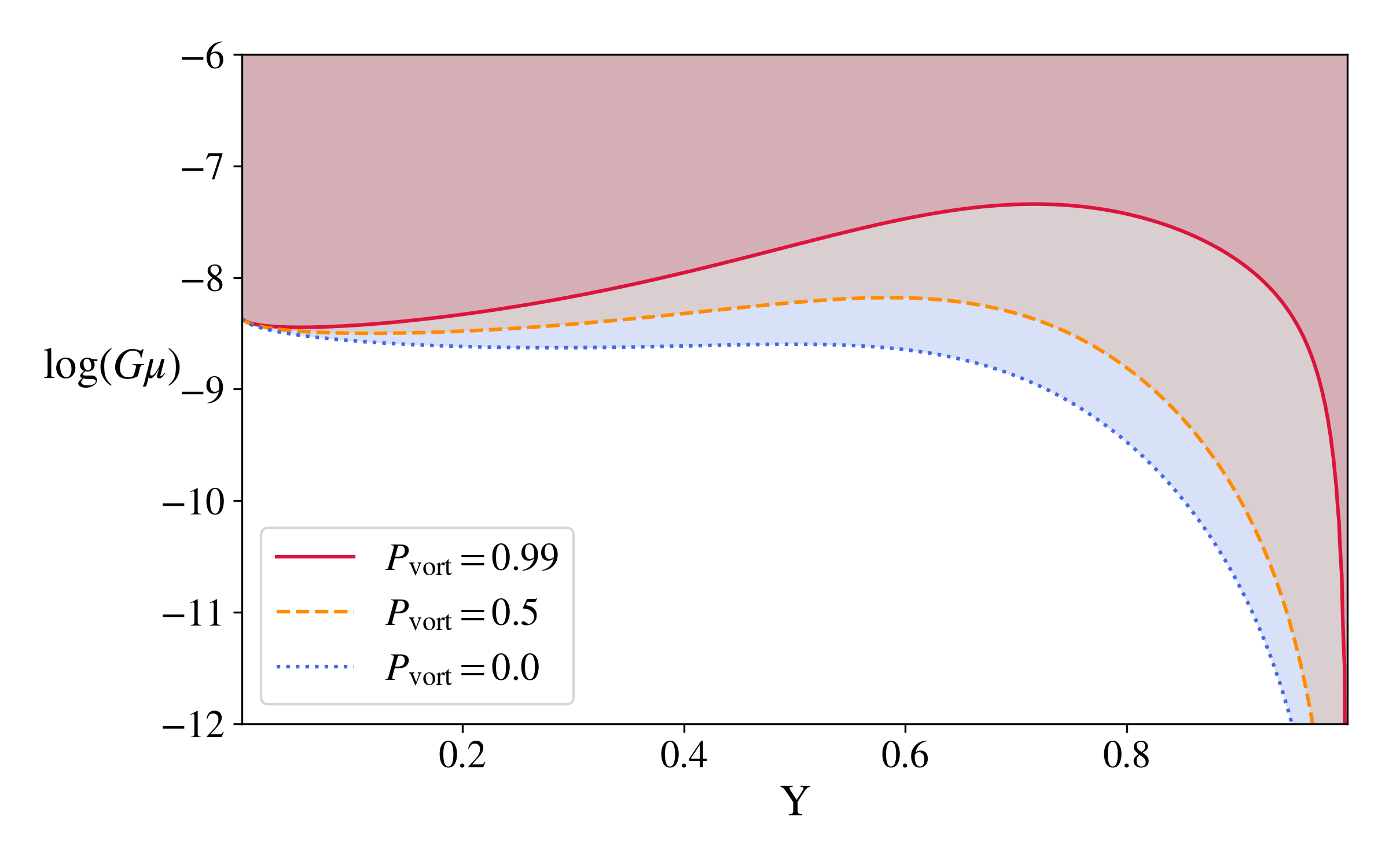}
\caption{\label{Fig.Constr} Constraints imposed by the LVK collaboration \cite{LIGOScientific:2025bgj} on cosmic string models producing a scale-invariant SGWB at $f=25 \,{\rm Hz}$, shown for different values of the current amplitude $Y$ and the vorton parameter $P_{\rm vort}$. }
\end{center}
\end{figure}

Gravitational wave observations provide additional constraints on cosmic string scenarios. We first analyze current-carrying string models using the NANOGrav 15-year PTA data \cite{NANOGrav:2023gor}. We use \texttt{PTArcade} \cite{Mitridate:2023oar,Lamb:2023jls} with the \texttt{enterprise} package \cite{2019ascl.soft12015E}, adopting $2 \times 10^6$ samples, $30$ red-noise components, and $14$ frequency bins, with model selection enabled. We consider the parameter ranges $\sqrt{\kappa_d} \in \left[7,9\right]$ and $\log(G\mu) \in \left[-10,-4\right]$, and, for current-carrying strings, $Y \in \left[0,0.99\right]$. We adopt uniform prior distributions for all these parameters. For standard strings, we fix the spectral index to $q=4/3$, corresponding to the standard prediction for cusp radiation, while for current-carrying strings, we set $q=5/3$, since cusp formation is suppressed in this scenario \cite{Rybak:2024our}. For strings without vortons, we fix $P_{\rm vort}=0$, while for strings with vortons, we set $P_{\rm vort}=1$ to investigate the maximal effect allowed by our phenomenological prescription. All the scenarios considered in this work, together with their parameter priors, are listed in Table \ref{tab:model_parameters}.

\begin{table}[t!]
\centering
\begin{tabular}{l||c|c|c|c|c|c}
\hline
Model &  $\log \alpha$ & $\log \mathcal{F}$ & $Y$ & $P_{\rm vort}$ & $\log e^2$ & $BF$ \\
\hline
\hline
Standard MSCS
    &  -1 & -1 & 0 & 0 & - & 1.0 \\
    \hline
MSCS with varying loop size 
    &  $\left[-3,  0 \right]$ & -1 & 0 & 0 & - & $1.6\pm0.4$ \\
    \hline
MSCS with varying $\mathcal{F}$ 
    &  -1 & $[-5, 0]$ & 0 & 0 & - & $0.4\pm0.1$ \\
    \hline
MSCS with current 
    &  -1 & -1 & $[0, 0.99]$ & 0 & - & $8.9\pm4.4$ \\
    \hline
MSCS with current and vortons 
    &  -1 & -1 & $[0, 0.99]$ & 1 & - & $1.1\pm0.4$ \\
    \hline
Superconducting MSCS 
    &  -1 & -1 & 0.1 & 0 &  $\left[-8, -1 \right]$ & $0.3\pm0.1$ \\
\hline
\end{tabular}
\caption{\label{tab:model_parameters} Prior distributions for the parameters of the MSCS models considered in this work. We also calculate the Bayes factors (BFs) of the alternative models relative to the standard MSCS model. For all models, we adopt uniform priors on $\log_{10}(G\mu)\in[-10,-4]$ and $\sqrt{\kappa_d}\in[7,9]$. The BFs are defined as the ratios of the Bayesian evidences of the alternative models to that of the standard MSCS model. }
\end{table}

\begin{figure}[t!]
\begin{center}
\includegraphics[scale=0.83]{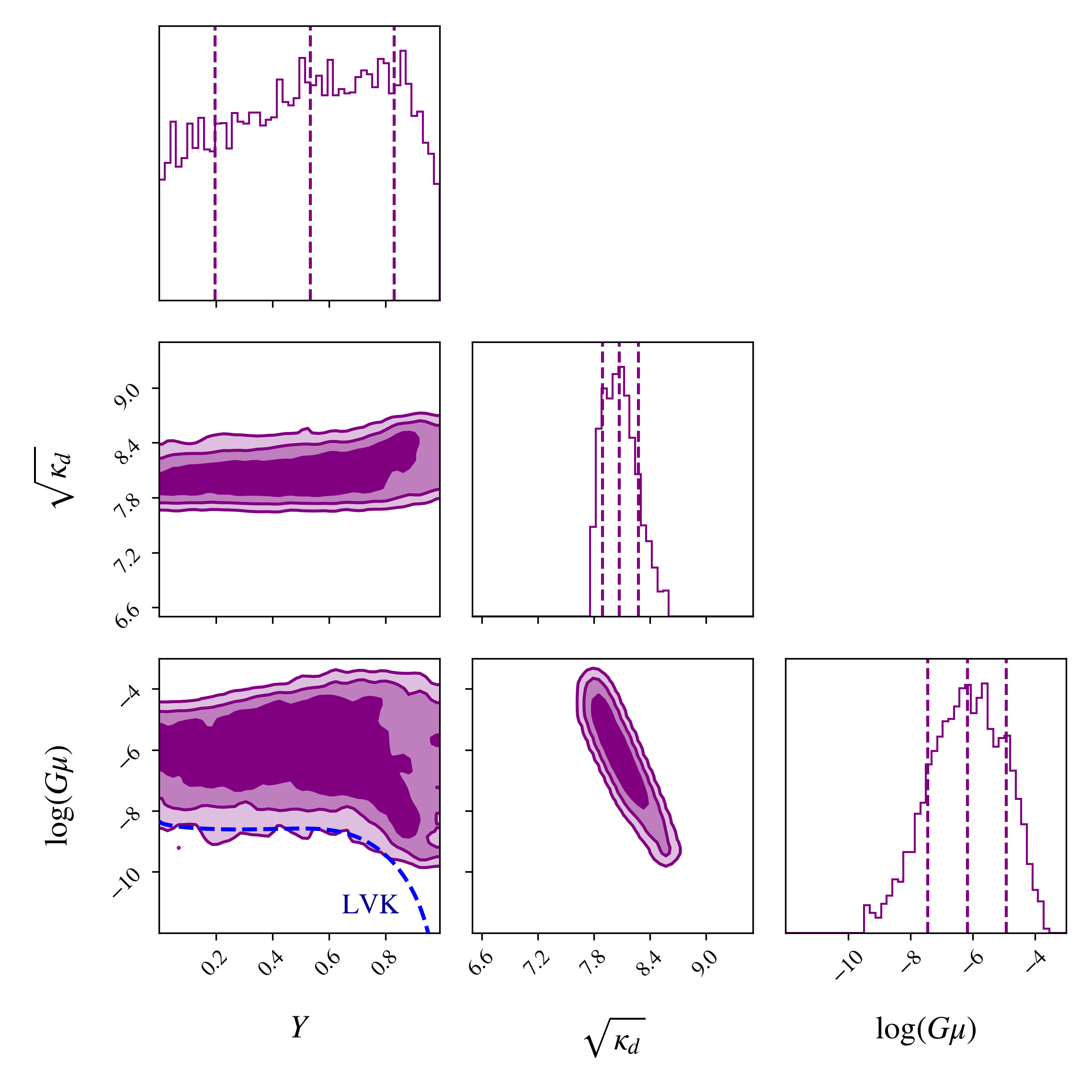}
\caption{\label{Fig.PTArcade_Y} Results of the analysis with \texttt{PTArcade} for current-carrying MSCSs, with uniform prior distributions $\sqrt{\kappa_d} \in \left[7,9\right]$, $\log(G\mu) \in \left[-10,-4\right]$ and $Y \in \left[ 0, 0.99 \right]$. The LVK constraint at 95\% credibility, $\Omega_{\rm gw}(25\,{\rm Hz}) < 2.8 \times 10^{-9}$ \cite{LIGOScientific:2025bgj}, is shown by the dashed line and is given by \eqref{LVK_Y_Constr}.  }
\end{center}
\end{figure}

From Table \ref{tab:model_parameters}, one can see that despite the introduction of the additional parameter $Y$, the BF of current-carrying strings is larger than that of standard metastable strings. This may be because almost the entire range of $Y$ is compatible with the PTA data. However, the differences between the BFs of standard MSCSs and the alternative models listed in Table \ref{tab:model_parameters} are not significant enough to claim an improvement in the fit over the standard MSCS model. In Fig.~\ref{Fig.PTArcade}, we show corner plots illustrating the parameter estimation for standard MSCSs. The results for current-carrying MSCSs without and with vortons are shown in Figs.~\ref{Fig.PTArcade_Y} and \ref{Fig.PTArcade_Y_P}, respectively. Corner plots showing variations in the other parameters can be found in Appendix~\ref{D}.

For a consistent analysis, we must also take into account the constraints from the LVK collaboration. In particular, we use the constraint on a scale-invariant SGWB,
\begin{equation}
\label{LVK_constr}
\Omega_{\rm gw}(25{\rm Hz}) < \Omega^{\rm LVK}_{\rm gw} = 2.8 \times 10^{-9},
\end{equation}
obtained in \cite{LIGOScientific:2025bgj}.  In the absence of modifications to the expansion history of the Universe, the MSCS spectra of all models considered here exhibit a scale-invariant plateau in the LVK frequency band for amplitudes satisfying \eqref{LVK_constr}. Therefore, we can directly apply the constraint in \eqref{LVK_constr} to probe cosmic string models. In particular, we use this bound to identify regions of parameter space compatible with the PTA signal while remaining consistent with the LVK constraints, and to assess whether metastable current-carrying cosmic strings can provide a viable explanation for the PTA signal.

To obtain the plateau contribution from cosmic strings, we follow the calculation in \cite{Sousa:2020sxs}. In particular, by generalizing Eq.~(22) of \cite{Sousa:2020sxs}, we obtain the lower limit of integration for current-carrying strings
\begin{equation}
a^{\rm plateau}_{\rm min} = \frac{4 H_0 \Omega_r^{1/2} S(Y)}{\Gamma_{\rm tot} \epsilon_r} \frac{j}{f}.
\end{equation}
Using $a^{\rm plateau}_{\rm min}$ as the lower integration limit and $a_{\rm vort}$, given by Eq.~\eqref{ScaleFactVort}, as the upper integration limit in Eq.~\eqref{OmegaR}, we find a broad frequency range over which the SGWB exhibits a plateau, corresponding to a scale-invariant SGWB. The amplitude of this plateau is given by
\begin{equation}
\label{PlateauR}
\Omega^{\rm plateau}_{\rm gw} = \frac{128 \pi}{9} \mathcal{F} A_r (G \mu)^2 \frac{\Gamma(Y)}{S(Y)} \Omega_r \Delta g_{i}^* \left( \left( 1 + \epsilon_r (1 - P_{\rm vort} Y_b ) \right)^{3/2} - 1 \right),
\end{equation}
where $A_r \equiv  c v \frac{\sqrt{1-v^2}}{\alpha \zeta^4}$. Here, $\Delta g_{i}^* \approx 0.387$ accounts for the change in the effective number of relativistic degrees of freedom according to \texttt{micrOMEGAs5.0} \cite{Belanger:2018ccd}. This value is appropriate for constraining $G\mu$ at a frequency of $25\,{\rm Hz}$, which corresponds to an epoch before the relevant change in the effective number of relativistic degrees of freedom.

\begin{figure}[t!]
\begin{center}
\includegraphics[scale=0.83]{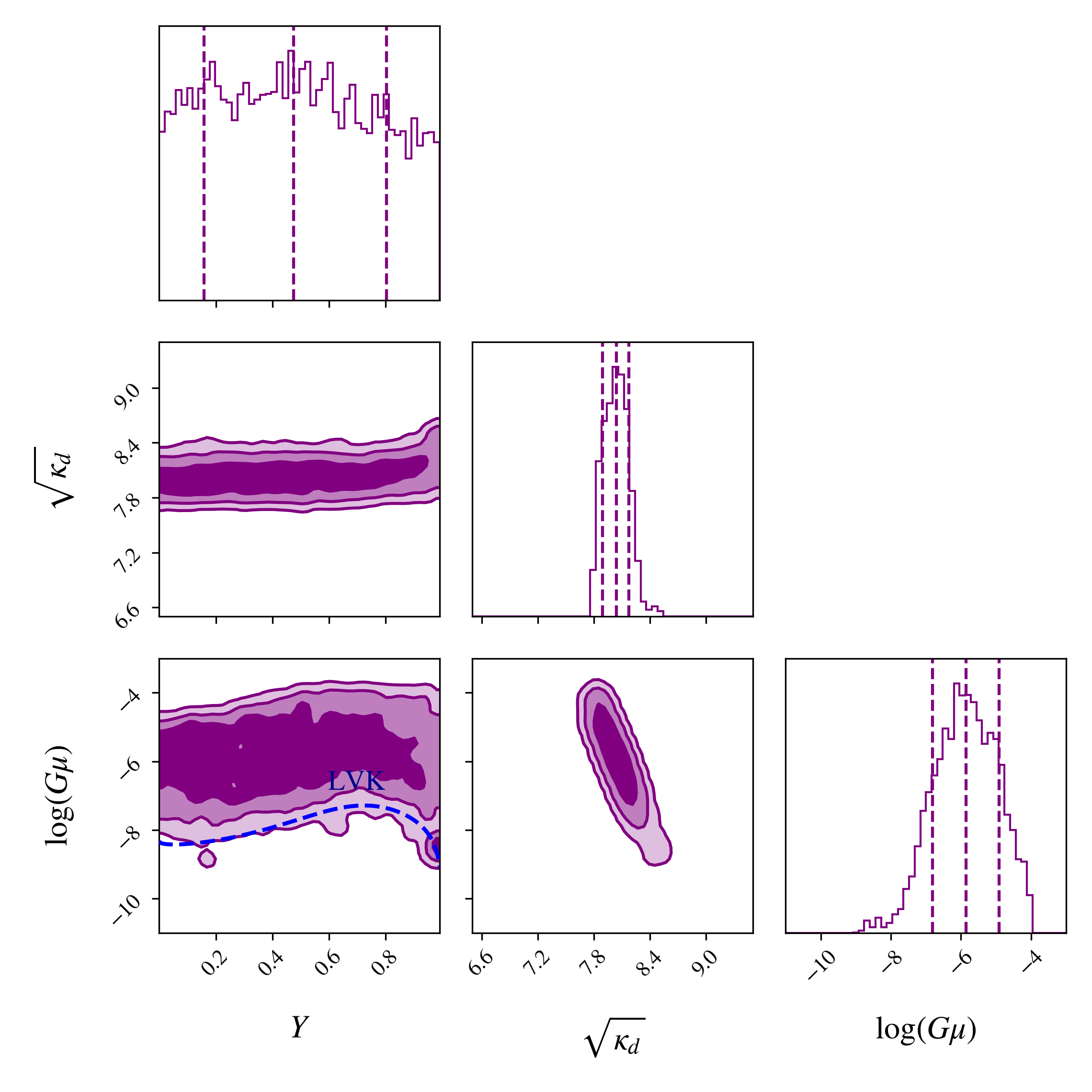}
\caption{\label{Fig.PTArcade_Y_P} Results of the analysis with \texttt{PTArcade} for current-carrying MSCSs with vortons $P_{\rm vort}=1$, with uniform prior distributions $\sqrt{\kappa_d} \in \left[7,9\right]$, $\log(G\mu) \in \left[-10,-4\right]$ and $Y \in \left[ 0, 0.99 \right]$. The LVK constraint at 95\% credibility, $\Omega_{\rm gw}(25\,{\rm Hz}) < 2.8 \times 10^{-9}$ \cite{LIGOScientific:2025bgj}, is shown by the dashed line and is given by \eqref{LVK_Y_Constr}.  }
\end{center}
\end{figure}

In the limit $P_{\rm vort} \rightarrow 0$, Eq.~\eqref{PlateauR} reduces to the result given by Eq.~(50) of \cite{Rybak:2024our}. Conversely, in the limit $P_{\rm vort} Y_b \rightarrow 1$, the SGWB approaches zero. This behavior is physically expected, since vorton formation prevents loops from decaying via gravitational radiation.

For strings without a current, $Y=0$, and assuming $\epsilon_r \gg 1$, we can further estimate the corresponding LVK constraint. At 95\% credibility, the standard MSCS scenario is constrained by
\begin{equation}
\label{LVK_std_Constr}
G\mu \lesssim \left( 24661 \times \Omega^{\rm LVK}_{\rm gw} \right)^2 \approx 4.8 \times 10^{-9}.
\end{equation}

For strings with a non-trivial current, the corresponding constraint depends on both $Y$ and $P_{\rm vort}$ and can be written as
\begin{equation} 
\label{LVK_Y_Constr}
G\mu \lesssim \left( \frac{ 632 \; \sqrt{\Gamma(Y_b)} \, S(Y_b) \, \Omega^{\rm LVK}_{\rm gw}}{(1-Y_b \, P_{\rm vort})^{3/2} \, \, (\alpha \zeta)^{3/2} \, A_r(Y_b) } \right)^2,
\end{equation}
where we have substituted the cosmological parameters and set $c=0.23$ and $\mathcal{F}=0.1$. The resulting constraints are illustrated in Fig.~\ref{Fig.Constr}.

When the LVK constraint is imposed on the parameter space obtained from the PTA analysis, we find that both the standard MSCS model and the current-carrying MSCS models that can account for the PTA signal are in tension with the LVK bound. Resolving this tension may require additional physics that suppresses the high-frequency SGWB or modifies the evolution of the string network. One possibility is a modified expansion history of the early Universe, which can alter the evolution of the string network and its resulting gravitational wave spectrum \cite{Cui:2017ufi,Datta:2024bqp}. Another possibility is an initial dilution or stretching of the string network during an early period of inflation \cite{Guedes:2018afo}, or the presence of friction that delays string network scaling \cite{Mukovnikov:2024zed, Hu:2025sxv, Mukovnikov:2026rpu}. Alternatively, the production of massive particles by small loops can suppress the high-frequency part of the SGWB and provide a high-frequency cutoff \cite{Auclair:2019jip}. However, in such scenarios, the resulting energy injection into the primordial plasma must also be consistent with the constraints of Big Bang nucleosynthesis \cite{SantanaMota:2014xpw, Kawasaki:2017bqm, Hindmarsh:2022awe}.

\begin{figure}[t!]
\begin{center}
\includegraphics[width=0.72\textwidth]{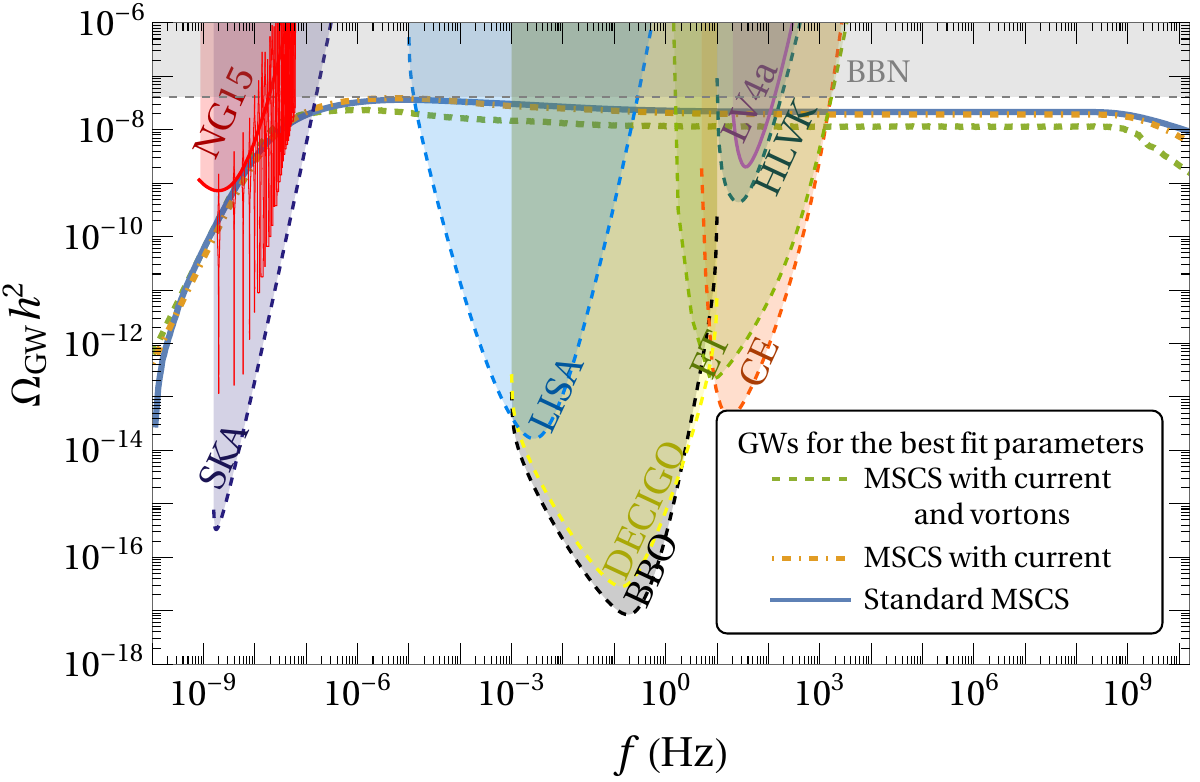}
\caption{\label{Fig.Omeg_fr} Spectra for the best-fit cases of standard and current-carrying MSCSs with and without vortons. The standard MSCS spectrum is represented by the solid line for $\log (G\mu) = -5.873$ and $\sqrt{\kappa_d} = 7.943$. The spectrum of current-carrying MSCSs without vortons is shown by the dash-dotted line for $\log (G\mu) = -6.230$, $\sqrt{\kappa_d} = 8.089$, and $Y=0.543$. The spectrum of current-carrying MSCSs with vortons is shown by the dashed line for $\log (G\mu) = -5.860$, $\sqrt{\kappa_d} = 8.037$, $Y=0.472$, and $P_{\rm vort}=1$. We also show the power-law integrated sensitivity curves \cite{Thrane:2013oya, Schmitz:2020syl} for several proposed experiments, namely SKA \cite{Janssen:2014dka}, LISA \cite{Bartolo:2016ami}, DECIGO \cite{Sato_2017}, BBO \cite{Crowder:2005nr, Corbin:2005ny}, HLVK \cite{KAGRA:2013rdx}, ET \cite{Mentasti:2020yyd}, and CE \cite{Regimbau:2016ike}, as well as bounds from big bang nucleosynthesis (BBN) \cite{Maggiore:1999vm}, LVK Run 4a (LV4a) \cite{LIGOScientific:2025bgj, LIGOScientific:2025kry}, and NANOGrav 15-year (NG15) data.}
\end{center}
\end{figure}

\begin{figure}[t!]
\begin{center}
\includegraphics[scale=0.34]{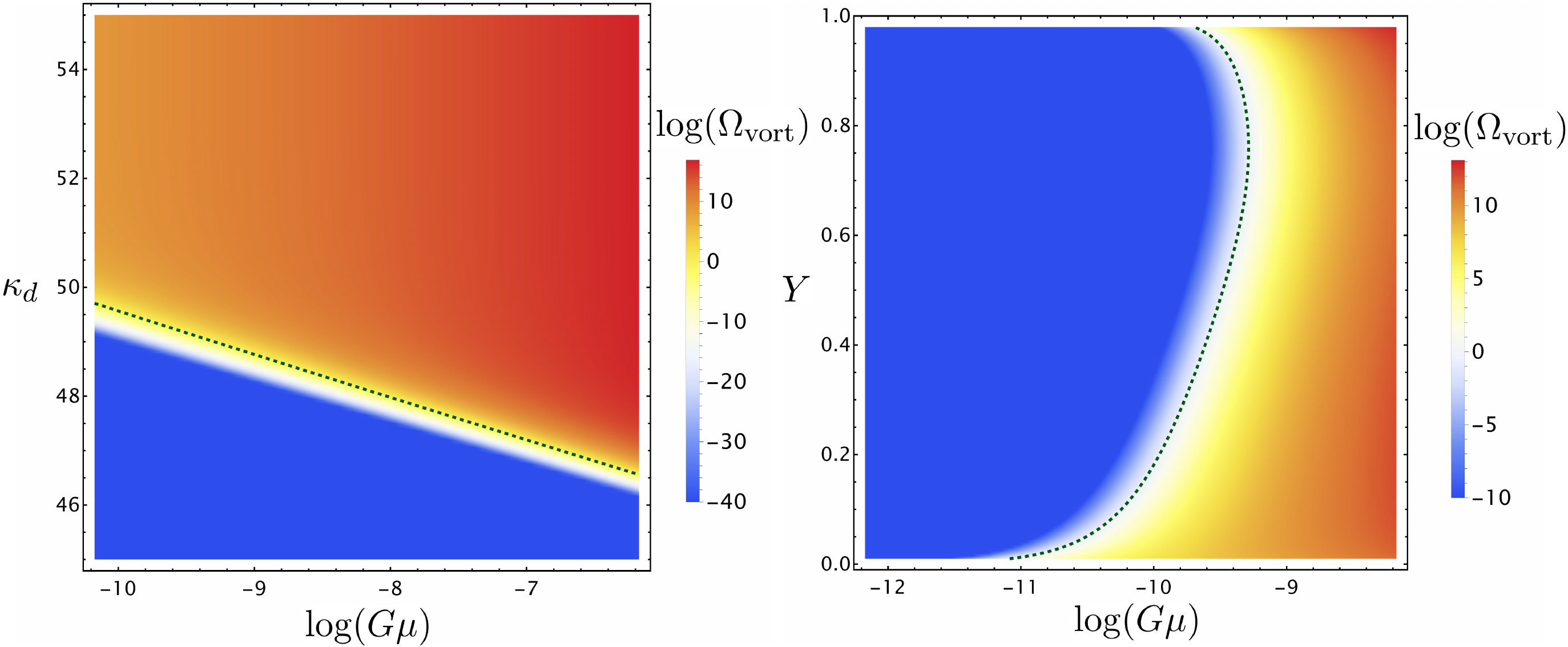}
\includegraphics[scale=0.34]{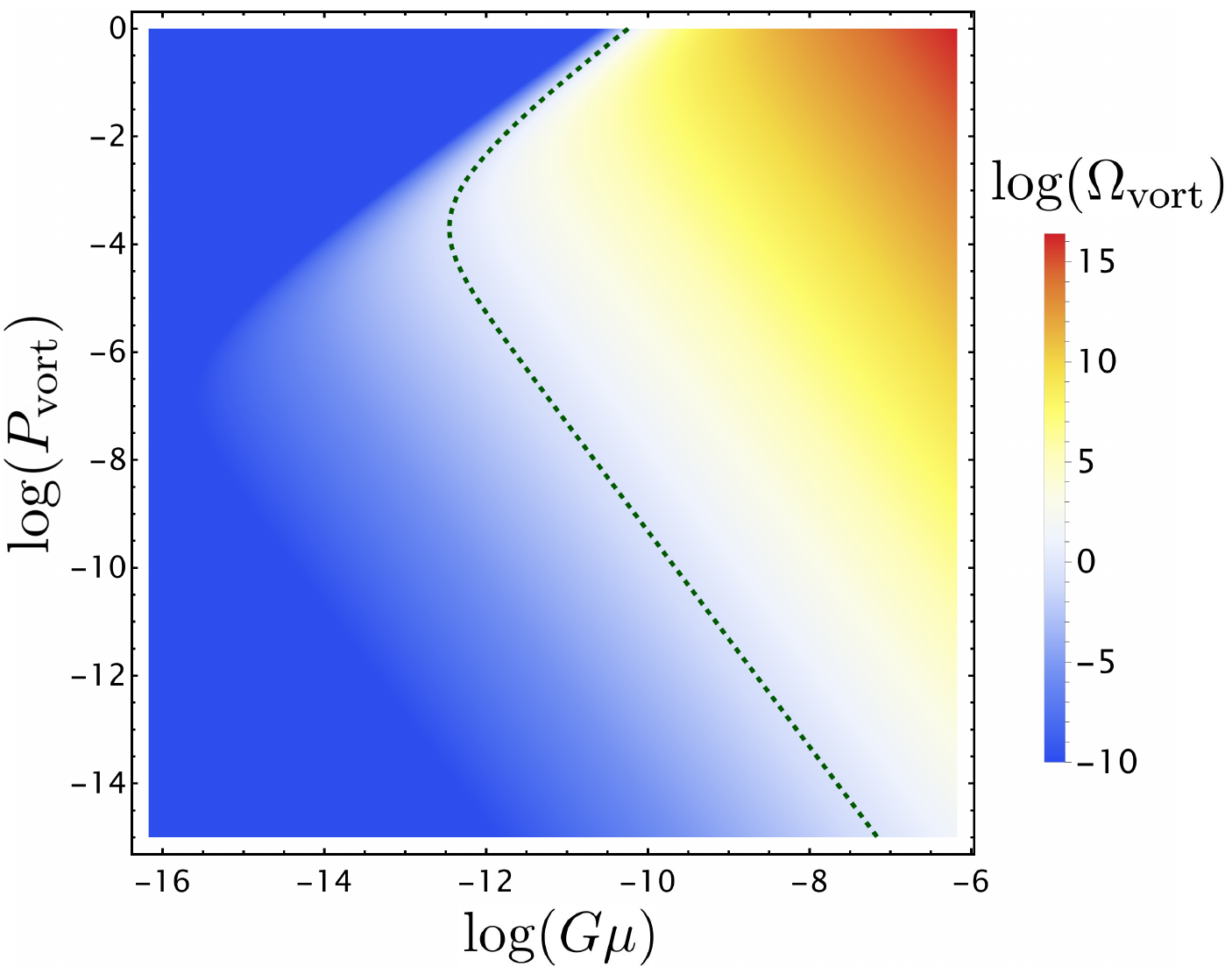}
\caption{\label{Fig.VortAb} Vorton abundance $\Omega_{\rm vort}$ for current-carrying MSCSs, as given by Eq.~\eqref{VortAbund}. The left upper panel shows the vorton abundance for $Y=0.8$ and $P_{\rm vort}=1$ for different values of $\kappa_d$ and $G\mu$. The right upper panel shows the vorton abundance for $\sqrt{\kappa_d}=7$ and $P_{\rm vort}=1$ for different values of $G\mu$ and $Y$. The lower panel shows the vorton abundance for fixed $Y=0.1$. }
\end{center}
\end{figure}

The MSCS model with vortons studied here demonstrates that the presence of vortons can provide an explanation for the PTA data while slightly reducing the tension with the LVK constraints; see Fig.~\ref{Fig.PTArcade_Y_P}. This scenario, however, requires a region of parameter space with $P_{\rm vort} \rightarrow 1$ and $Y \rightarrow 1$. For nonzero $P_{\rm vort}$, the vorton-abundance constraint must also be imposed \cite{Carter:1999an, Auclair:2020wse}. We impose the requirement that the energy density in vortons should not exceed the observed dark matter abundance, $\Omega_{\rm vort} \leq \Omega_{\rm DM} = 0.264$ \cite{Planck:2018vyg}. The details of our derivation of the vorton abundance constraint are given in Appendix~\ref{Vorton abundance}.

In deriving this constraint, we assume that the current condenses onto cosmic strings at approximately the time of string network formation. This assumption suppresses the SGWB plateau in the LVK frequency band, improving the agreement between the PTA and LVK observations. However, it also results in a vorton abundance that exceeds the dark matter abundance bound for all relevant parameter values, as shown in Fig.~\ref{Fig.VortAb}.

\section{Conclusions}
\label{VI}

In this work, we investigated the SGWB generated by metastable current-carrying cosmic strings. We obtained the loop production function within the CVOS description of the long-string network. We also extended the analysis to include the probability of monopole-antimonopole nucleation and a phenomenological prescription for vorton formation. This framework allows us to study how additional degrees of freedom on the string worldsheet can modify the gravitational wave spectrum relative to the standard NG description.

In this study, we restricted the analysis to the chiral current, for which the monopole-antimonopole nucleation rate is the same as that for ordinary metastable strings. At the same time, the presence of the current modifies the loop dynamics, the gravitational wave emission efficiency, and the relation between the loop length and the emitted radiation frequency. The presence of a current changes both the amplitude and the characteristic frequencies of the SGWB. Vector radiation can provide an additional channel for loop energy loss, which can suppress the gravitational wave signal. Vorton formation has a different effect: within our phenomenological prescription, it mainly modifies the high-frequency part of the spectrum, while the low-frequency spectrum is comparatively insensitive to vorton formation.

An important result of our analysis is the identification of two different low-frequency regimes. If the exponential suppression of the loop distribution determines the upper limit of the gravitational wave integral, the spectrum approaches an $f^2$ behavior, as shown in \cite{Buchmuller:2021mbb}. In contrast, if the finite lifetime of the metastable string network determines the relevant cutoff, the asymptotic behavior becomes $\Omega_{\rm gw}\propto f^3$. The latter therefore arises naturally and does not require an additional phenomenological separation of timescales, as introduced in \cite{Asl:2026zpj}. Depending on the model parameters, the $f^3$ regime may occur at sufficiently low frequencies, with the $f^2$ behavior appearing as an intermediate or transient regime.

We also constructed an analytical template for the SGWB that captures the main features of the full numerical calculation. The template incorporates the relevant cutoff scales, the evolution of the relativistic degrees of freedom, and the phenomenological vorton cutoff. In the PTA frequency range, the dominant contribution comes from the lowest modes of the largest loops. This allows us to use an explicit summation over the first $10^3-10^4$ modes, which provides good agreement with the full numerical calculation.

We used this fast template to probe current-carrying MSCSs with the NANOGrav 15-year data. The current-carrying models yield BFs comparable to those of the corresponding standard MSCS model (see Table \ref{tab:model_parameters}). The result shows that current-carrying MSCSs can provide a comparably good phenomenological description of the PTA signal while allowing for additional structure in the spectrum.

We then considered the constraint from the LVK search for a scale-invariant SGWB. Although regions of parameter space can reproduce the PTA signal, both the standard MSCSs and the current-carrying extensions considered here remain in tension with the LVK upper bound or violate the vorton abundance constraints. The inclusion of a worldsheet current and vortons alone does not resolve the tension between explaining the PTA signal and meeting the LVK and vorton-abundance constraints in MSCS models. Additional physics capable of suppressing the high-frequency gravitational wave spectrum may therefore be required. We also note that our analysis does not take into account the contribution from cosmic string segments \cite{Chitose:2025qyt}.

Our results demonstrate that currents on cosmic strings can lead to qualitatively new features in their SGWB. The analytical template developed here provides a practical framework for exploring these effects using ongoing PTA, LVK, and future gravitational-wave observations, which may distinguish current-carrying from standard MSCSs.

\section*{Acknowledgements}
 I.R. and R.M. are supported by IBS under the project code IBS-R018-D3.
I.R. would also like to thank Sergei Mukovnikov for useful discussions on various aspects of this work, the organizers and participants of the workshop ``Universe Unravelled: Cosmic Strings at 50'' for the opportunity to present these results, and Ken D. Olum for his critical comments.

\appendix
\setcounter{equation}{0}
\renewcommand{\theequation}{A.\arabic{equation}}

\section{Monopole-antimonopole nucleation for current-carrying cosmic strings}
\label{Monopole-antimonopole nucleation for current-carrying cosmic strings}

To estimate the decay rate of current-carrying cosmic strings via monopole-antimonopole pair nucleation, we follow the semiclassical approach of Ref.~\cite{Preskill:1992ck}, which builds upon the original analysis of Ref.~\cite{Vilenkin:1982hm}. 
Consider a sufficiently small segment of a cosmic string, which can be approximated locally as a straight string aligned with the $z$-axis. We further assume that the characteristic radius of the Euclidean bounce is much smaller than the length scale over which the worldsheet current varies. As a result, the equation of state can be evaluated locally, and the state parameter may be approximated by a constant value, $\chi \simeq \chi_{\rm seg}$, throughout the tunneling process. Under this approximation, the master function $\mathcal{L}(\chi)$ may be treated as a constant.

After performing a Wick rotation for the action \eqref{Action}, the Euclidean action describing a string with monopole endpoints becomes
\begin{equation}
S_E = m\int d\lambda + \mu\int \mathcal{L}(\chi)\sqrt{\gamma} d^2\zeta,
\end{equation}
where $m$ is the monopole mass and $\lambda$ parametrizes the monopole worldline. Since the interaction between the current carriers and the monopole is not yet fully understood (see, e.g., Refs.~\cite{Hindmarsh:1989jq, Holman:1992rv}), we assume that the monopole properties, in particular its mass, are unaffected by the current of the string. Within these assumptions, the Euclidean action differs from the ordinary MSCS case only through the replacement $\mu \rightarrow \mu \mathcal{L}(\chi_{\rm seg})$. 

 Following Ref.~\cite{Preskill:1992ck}, the bounce configuration corresponds to a nucleating circular hole in the Euclidean worldsheet of area $\pi R^2$, while the monopole and antimonopole trace worldlines of total length $2\pi R$. The corresponding bounce action is therefore
\begin{equation}
B(R)=2\pi Rm-\pi R^2 \mu \mathcal{L}(\chi_{\rm seg}).
\end{equation}
Extremizing the action with respect to the bounce radius gives
\begin{equation}
R=\frac{m}{\mu \mathcal{L}(\chi_{\rm seg})},
\end{equation}
and the on-shell bounce action becomes
\begin{equation}
B=\pi\frac{m^2}{\mu \mathcal{L}(\chi_{\rm seg})}.
\end{equation}
The corresponding semiclassical decay rate is then exponentially suppressed according to \cite{LifshitzKagan, Coleman:1977py, Coleman:1987rm}
\begin{equation}
\label{ProbabilityA}
\Gamma_d \approx \frac{\mu \mathcal{L}(\chi_{\rm seg})}{2 \pi}
\exp\left[- \pi \kappa_d \right],
\end{equation}
where $\kappa_d \equiv  \frac{ m^2}{\mu \mathcal{L}(\chi_{\rm seg})}$ and the calculation of the prefactor can be carried out in the same manner as in Refs.~\cite{Kiselev:1975eq, Monin:2008mp}.

Eq.~\eqref{ProbabilityA} shows that the presence of a worldsheet current can modify the decay rate through the master function $\mathcal{L}(\chi)$ which enters the tunneling exponent. For example, in Witten's $U(1)\times U(1)$ superconducting string model \cite{Witten:1984eb}, the equation of state is encoded in the master function \cite{Peter:1992dw,Peter:1992ta,Carter:1994hn}
\begin{equation}
\mathcal{L}(\chi)=
\begin{cases}
1+\dfrac{\ln(1-2\alpha_{\sigma} \chi)}{4 \alpha_{\sigma} }, \quad \chi\geq 0 \\
1-\dfrac{1}{2}\dfrac{\chi}{1-\alpha_{\sigma} \chi},
\quad \chi\leq 0,
\end{cases}
\end{equation}
where $\alpha_{\sigma} \equiv \frac{m_H^2}{m_\sigma^2}$ is the ratio of the squared Higgs mass associated with string formation to the squared mass of the current-carrying condensate. Other microscopic models lead to different forms of the master function and therefore may potentially predict different tunneling rates \cite{Carter:1999hx,Carter:2003fb,Carter:1989dp,Hartmann:2008yr}.

The above argument demonstrates that worldsheet currents can, in principle, alter the metastability of cosmic strings by modifying the monopole-antimonopole nucleation rate. This effect should therefore be taken into account when constructing realistic models of metastable superconducting cosmic strings. Although the worldsheet current can generically modify the monopole-antimonopole nucleation rate, here we consider the chiral current, i.e., $\chi \rightarrow 0$, for which the master function satisfies $\mathcal{L}(0)=1$. Hence, the metastability parameter $\kappa_d$ is independent of the current in the present model.

A more accurate determination of the decay rate requires a full field-theoretic treatment of the nucleation process, as discussed in Refs.~\cite{Shifman:2002yi,Chitose:2023dam}, together with the inclusion of finite-temperature corrections \cite{Holman:1992rv,Monin:2008uj,Tranchedone:2026lav}. Such an analysis is beyond the scope of this paper and is left for future investigation.

\section{Loop distribution function for metastable current-carrying cosmic string networks}
\label{Loop distribution function for metastable current-carrying cosmic string network}

\setcounter{equation}{0}
\renewcommand{\theequation}{B.\arabic{equation}}

We consider the equations of motion for an individual cosmic string carrying a light-like (chiral) current, which follow from varying the action~\eqref{Action}. The corresponding equations of motion are given by \cite{Martins:2020jbq}
\begin{equation}
\begin{gathered}
\label{LoopEqOfMot}
    \partial_{\tau} \left[ \varepsilon (1+y) \right] + 2 \frac{\dot{a}}{a} \varepsilon \left[  \dot{\textbf{X}}^2 + y \right] = y^{\prime}, \\
    \ddot{\textbf{X}} \varepsilon (1+y) + 2 \frac{\dot{a}}{a} \varepsilon \left( 1 - \dot{\textbf{X}}^2 \right) \dot{\textbf{X}} = \left( \frac{1-y}{\varepsilon} \textbf{X}^{\prime}  \right)^{\prime} + 2 y \dot{\textbf{X}}^{\prime} + \textbf{X}^{\prime} \left( \dot{y} + 2 \frac{\dot{a}}{a} y \right), \\
    \partial_{\tau} \left[ a \sqrt{y} \sqrt{\textbf{X}^{\prime \, 2}}  \right] = \partial_{\sigma} \left[ a \sqrt{y} \sqrt{1 - \dot{\textbf{X}}}^2 \right],
\end{gathered}
\end{equation}
where $\varepsilon^2 \equiv \frac{\textbf{X}^{\prime \, 2}}{1 - \dot{\textbf{X}}^2}$, $y \equiv - 2 \gamma^{00} \dot{\varphi}^2 f_{\chi} $, following the notation of Ref.~\cite{Martins:2020jbq}. In the chiral limit, $\chi \rightarrow 0$, the derivative of the master function becomes a constant, $f_\chi=\mathrm{const.}$, while the quantity $y$ remains a dynamical variable through its dependence on the worldsheet field $\varphi$, see Ref.~\cite{Martins:2020jbq} for details.
The total energy of a loop is given by
\begin{equation}
\label{EnergyLoops}
    E = \mu a \int \varepsilon \left( 1 + y \right) d \sigma.
\end{equation}
Integrating Eq.~\eqref{LoopEqOfMot} over the worldsheet coordinate $\sigma$ and summing over all loops yields the evolution equation for the loop energy density. Defining $\rho_L=E/V_h$ and using Eq.~\eqref{EnergyLoops}, one obtains
\begin{equation}
    \frac{d \rho_L}{d t} = - 2 H \rho_L \left( 1 + \left\langle \frac{\dot{\textbf{X}}^2 + y }{1 + y} \right\rangle \right) + \left( 1 + Y \right) \mu l f_L(l, t) .
\end{equation}
Here, the macroscopic variable $Y$ characterizes the long strings at the time of loop production and should not be confused with the variable $y$, which depends on $(\sigma,\tau)$ for an individual loop. The function $f_L(l,t)$ denotes the loop production function, and $l$ is the invariant loop length. It is convenient to express the loop energy density in terms of the loop number density $n(l,t)$ as
\begin{equation}
\label{LoopNumberDens1}
\rho_L(l, t) \equiv \mu (1+Y_L) l n(l,t),
\end{equation}
where $Y_L \equiv \left\langle y \right \rangle$ denotes the average charge carried by the loop, while $Y$ is the charge of long strings.

In the case of MSCSs, the loop energy density acquires an additional sink term describing loop decay with probability per unit length $\Gamma_d$. Since the decay rate is proportional to the loop length, the evolution equation becomes
\begin{equation}
\label{LoopDensEq}
    \frac{d \rho_L}{d t} = - 2 H \rho_L \left( 1 + \left\langle \frac{\dot{\textbf{X}}^2 + y }{ 1 + y} \right\rangle \right) + \left( 1 + Y \right) \mu  l f_L(l,t) - \Gamma_d l \rho_L.
\end{equation}
If we assume that the oscillation period of a loop is much shorter than the Hubble time, so that the effects of the expansion of the Universe can be averaged over one oscillation period, then one can show that $\left\langle \frac{\dot{\textbf{X}}^2+y}{1+y} \right\rangle = 1/2$. Under this approximation, current-carrying cosmic string loops behave as non-relativistic matter with a source function $f_L(l,t)$ and decay channel $\Gamma_d$:
\begin{equation}
\label{LoopDensEq_small}
    \frac{d \rho_L}{d t} = - 3 H \rho_L + \left( 1 + Y \right) \mu  l f_L(l,t) - \Gamma_d l \rho_L.
\end{equation}
Equation~\eqref{LoopDensEq_small} can be rewritten as
\begin{equation}
\label{LoopDensEq2G}
    \frac{d}{d t} \left( \rho_L a^3 \text{e}^{\int_{t_i}^t \Gamma_d l dt' } \right) =  a^3 \text{e}^{\int_{t_i}^t \Gamma_d l dt' }  \left( 1 + Y \right) \mu  l f_L(l,t) ,
\end{equation}
where we assume that no loops are present at the initial time, namely $\rho_L(l(t_i), t_i) = 0$. The solution is therefore
\begin{equation}
\label{LoopDensEq3G}
     \rho_L(l,t) = \int^{t}_{t_i} \mu l \left( \frac{a(t')}{a(t)} \right)^3 \left( 1 + Y(t') \right) \text{e}^{- \int_{t'}^{t} \Gamma_d l(t'') dt'' }    f_L(l(t'),t') dt' .
\end{equation}
Using Eq.~\eqref{LoopNumberDens1}, we obtain the loop number density
\begin{equation}
    \label{LoopNumbDensG}
    n(l(t),t) = \int_{t{_i}}^t \frac{1+Y(t')}{1+Y_L(t)} \left( \frac{a(t')}{a(t)} \right)^3  \text{e}^{- \int_{t'}^{t} \Gamma_d l(t'') dt'' } f_L (l(t'),t') dt'.
\end{equation}
The loop production function $f_L(l,t)$ is related to the chopping efficiency through
\begin{equation}
   c \rho \frac{v}{\xi} = \mu \left( 1 + Y \right)\int_0^{\infty} l f_L (l,t) d l. 
\end{equation}
Using $\rho=\mu/L^2$, this expression can be rewritten as
\begin{equation}
\label{LoopConnect}
   c \frac{v}{\xi^3} =  \int_0^{\infty} l f_L (l) d l, 
\end{equation}
where we have used the relation $\xi^2=(1+Y)L^2$ (see Eq.~(33) of \cite{Martins:2020jbq}). We assume that all loops are produced with the same initial size, corresponding to a fixed fraction of the characteristic length, $l=\alpha\xi$, which can be modeled by a delta-function distribution. Using Eq.~\eqref{LoopConnect}, the loop production function becomes
\begin{equation}
\label{LoopProductFunct}
    f_L (l,t) = \mathcal{F} \sqrt{1-v^2}  \frac{c v}{l \xi^3} \delta(l-\alpha \xi),
\end{equation}
where $\mathcal F$ is a fuzziness parameter that specifies the fraction of loops contributing to the SGWB. The Lorentz factor accounts for the fact that newly produced loops possess center-of-mass kinetic energy, which is subsequently redshifted by the expansion of the Universe rather than being emitted as gravitational waves.

Substituting Eqs.~\eqref{LoopProductFunct} and \eqref{LengthChange} into Eq.~\eqref{LoopNumbDensG}, we obtain
\begin{equation}
    n(l(t),t) = \mathcal{F} \sqrt{1-v_b^2} \frac{c v_b}{\alpha \xi_b^4} \left( \frac{a_b}{a(t)} \right)^3   \frac{ \exp \left[- \Gamma_d \left( l ( t - t_b ) + \Gamma_{\rm tot} \frac{\left( t - t_b \right)^2}{2}  \right) \right] }{ \alpha \frac{d \xi}{dt} \big|_{t=t_{b}} + \Gamma_{\rm tot} } ,
\end{equation}
where we have assumed that $Y_L(t)=Y(t_b)$, following Ref.~\cite{Rybak:2024our}. All quantities carrying the subscript $b$ are evaluated at the loop formation time $t_b$, and the expression coincides with the one given in the main text \eqref{LoopNumbDensG2}. 

\begin{figure}[t!]
\begin{center}
\includegraphics[scale=0.83]{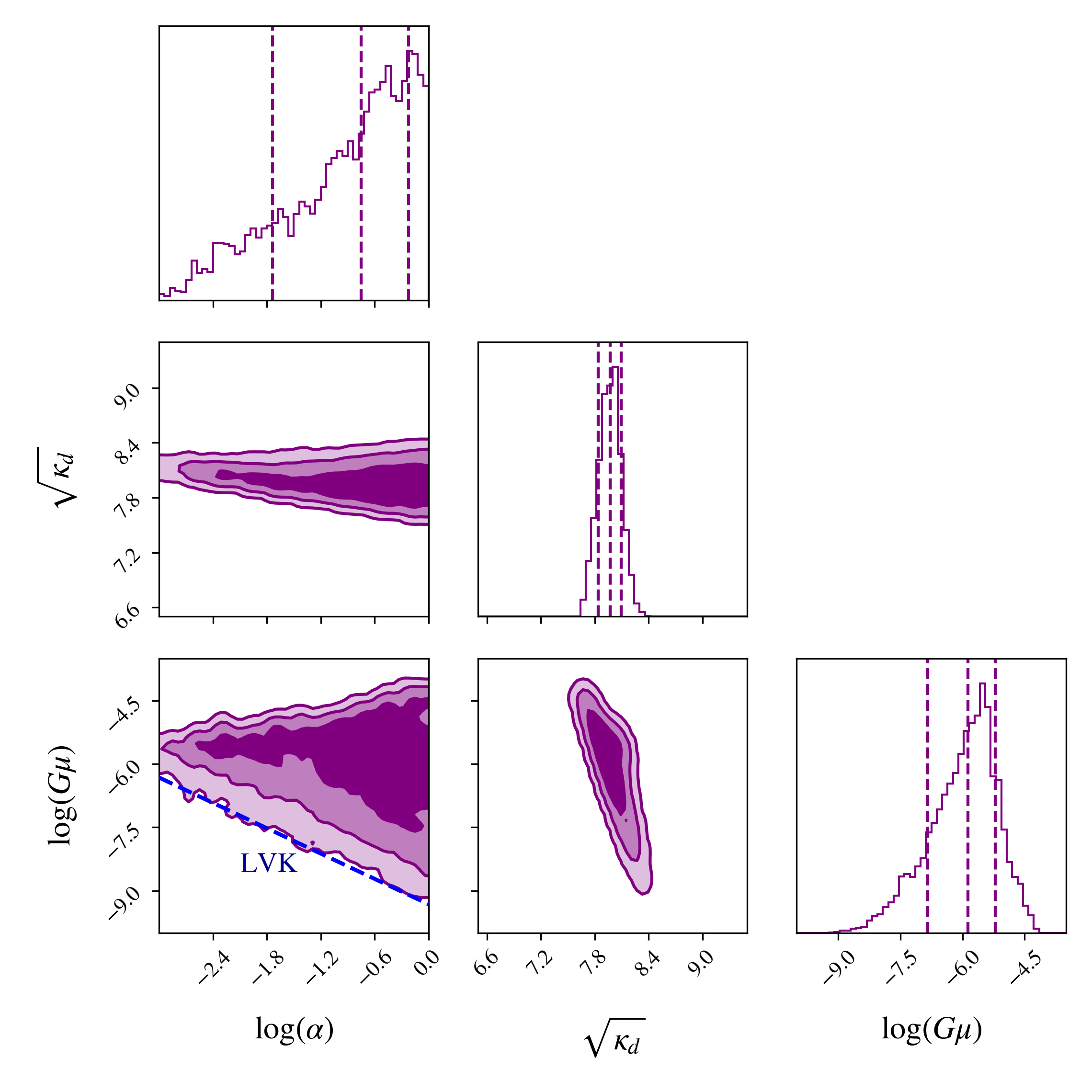}
\caption{\label{Fig.PTArcade_alpha}  Results of the analysis with \texttt{PTArcade} for MSCSs with uniform prior distributions $\sqrt{\kappa_d} \in \left[7,9\right]$, $\log(G\mu) \in \left[-10,-4\right]$ and $\log(\alpha) \in \left[ -3, 0 \right]$. The LVK constraint at 95\% credibility, $\Omega_{\rm gw}(25\,{\rm Hz}) < 2.8 \times 10^{-9}$ \cite{LIGOScientific:2025bgj}, is shown by the dashed line and is given by \eqref{LVK_Y_Constr}.  }
\end{center}
\end{figure}

\begin{figure}[t!]
\begin{center}
\includegraphics[scale=0.83]{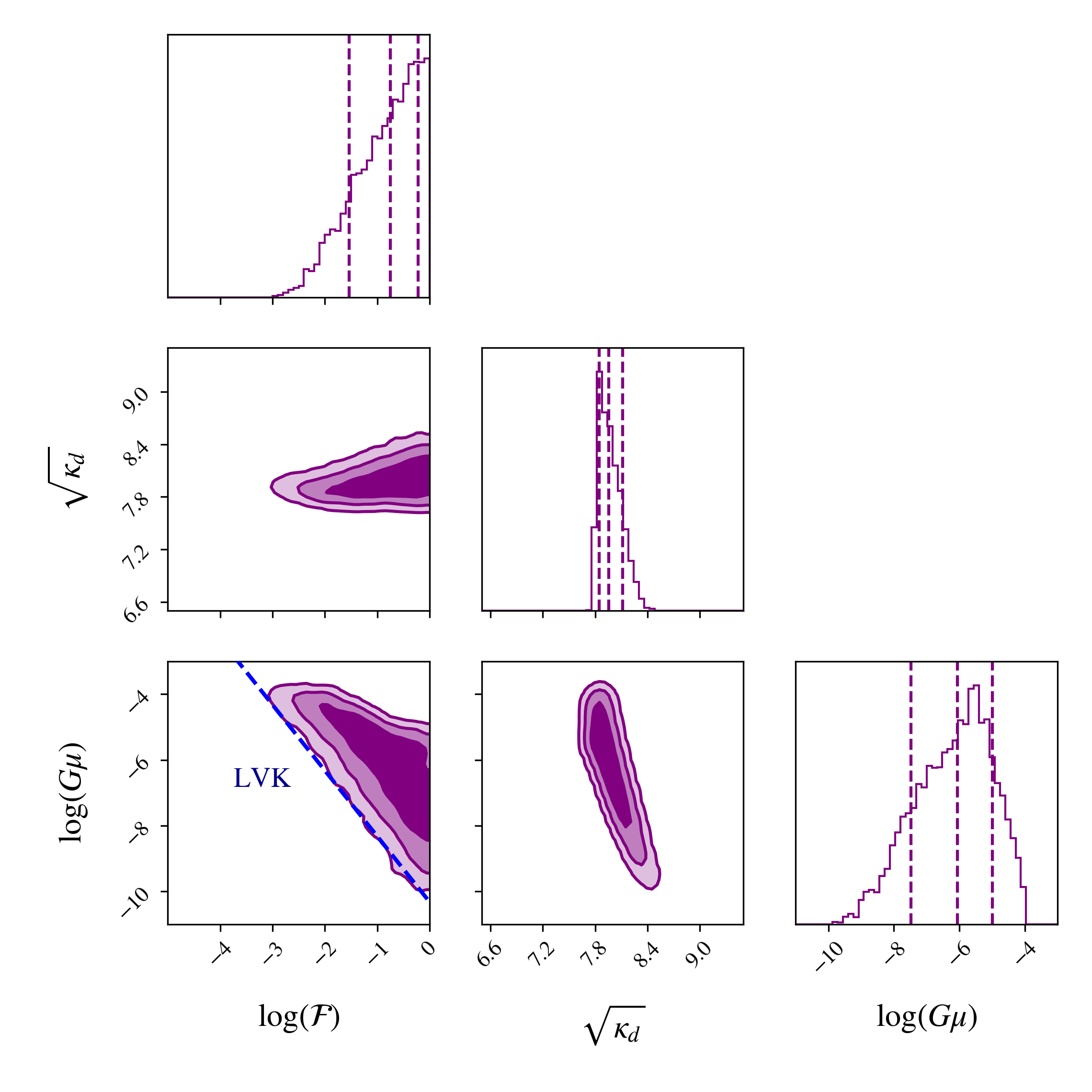}
\caption{\label{Fig.PTArcade_F} Results of the analysis with \texttt{PTArcade} for MSCSs with uniform prior distributions $\sqrt{\kappa_d} \in \left[7,9\right]$, $\log(G\mu) \in \left[-10,-4\right]$ and $\log(\mathcal{F}) \in \left[ -5, 0 \right]$. The LVK constraint at 95\% credibility, $\Omega_{\rm gw}(25\,{\rm Hz}) < 2.8 \times 10^{-9}$ \cite{LIGOScientific:2025bgj}, is shown by the dashed line and is given by \eqref{LVK_Y_Constr}.}
\end{center}
\end{figure}

\begin{figure}[t!]
\begin{center}
\includegraphics[scale=0.83]{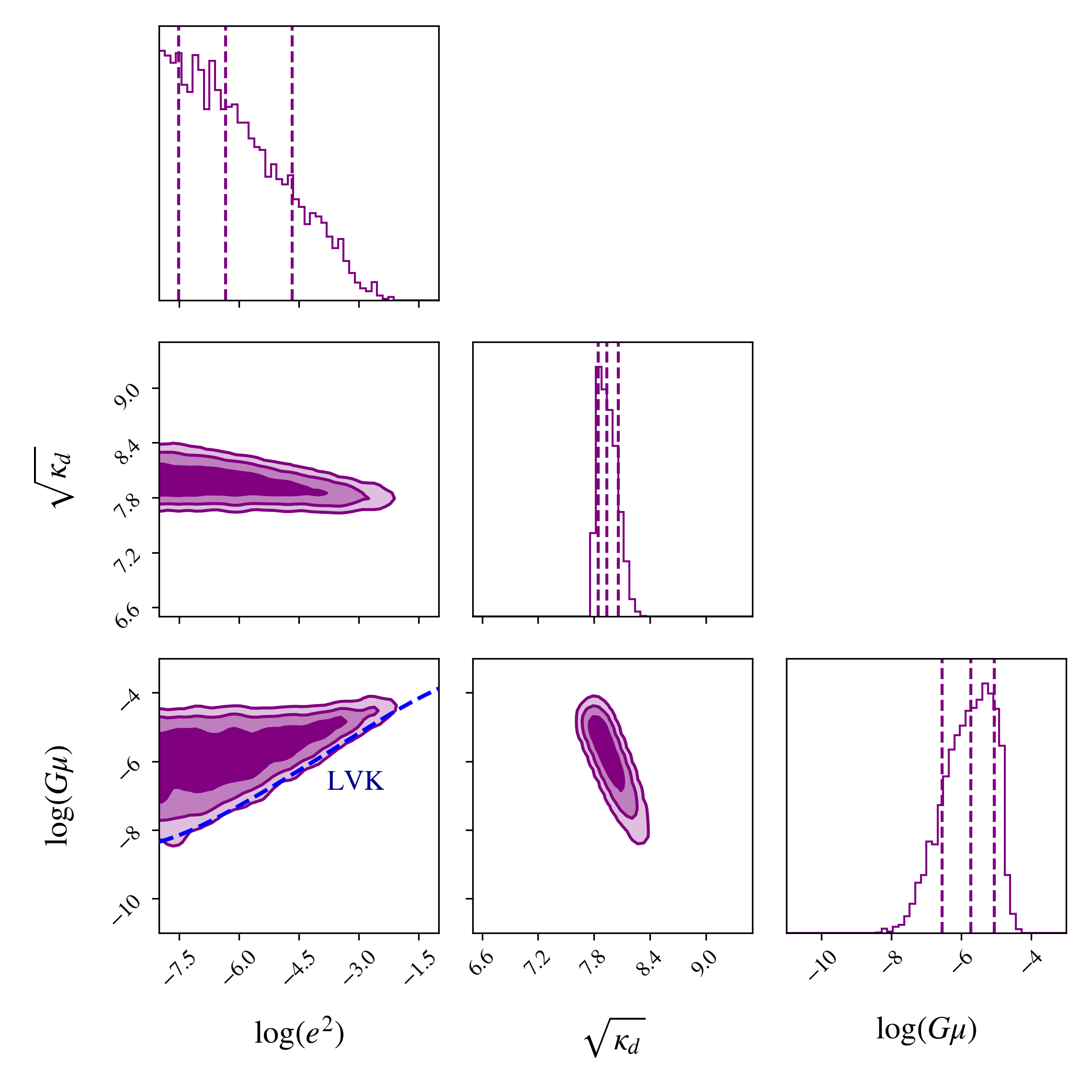}
\caption{\label{Fig.PTArcade_e} Results of the analysis with \texttt{PTArcade} for MSCSs with a current and nonzero coupling $e^2$ to the associated field. We adopt uniform priors on $\sqrt{\kappa_d} \in \left[7,9\right]$, $\log(G\mu) \in \left[-10,-4\right]$, and $\log(e^2) \in \left[ -8, -1 \right]$ with $Y$ fixed to $0.1$. The LVK constraint at 95\% credibility, $\Omega_{\rm gw}(25\,{\rm Hz}) < 2.8 \times 10^{-9}$ \cite{LIGOScientific:2025bgj}, is shown by the dashed line and is given by \eqref{LVK_Y_Constr}.  }
\end{center}
\end{figure}

\section{Vorton abundance}
\label{Vorton abundance}

\setcounter{equation}{0}
\renewcommand{\theequation}{C.\arabic{equation}}

Using the results of Appendix \ref{Loop distribution function for metastable current-carrying cosmic string network}, we can write the evolution of the vorton energy density as
\begin{equation}
\label{LoopDensEq_smallVort}
    \frac{d \rho_V}{d t} = - 3 H \rho_V + \left( 1 + Y \right) \mu  l_V f_V(l,t) - \Gamma_d l_V \rho_V.
\end{equation}
The solution to this equation is given by
\begin{equation}
\label{VortDensEq3G}
     \rho_V(l_V,t) = \int^{t}_{t_i} \mu \left( \frac{a(t')}{a(t)} \right)^3 \left( 1 + Y(t') \right) \text{e}^{- \int_{t'}^{t} \Gamma_d l_V (t'') dt'' }  l_V  f_V(l(t'),t') dt' ,
\end{equation}
where the vorton length $l_V$ is fixed, in contrast to the length of loops, since vorton loops do not shrink. The corresponding vorton production function is
\begin{equation}
\label{VortonProductFunct}
    f_V (l_V,t) = \mathcal{F} P_{\rm vort} Y_b \sqrt{1-v^2}  \frac{c v}{l_V \xi^3} \delta \left( l_V-\alpha \zeta_b t_b Y_b P_{\rm vort} \right).
\end{equation}
Integrating over $l_V$, we obtain the present-day vorton energy density. Dividing this result by the critical density of the Universe, we obtain the vorton energy-density fraction
\begin{equation}
\label{VortAbund}
    \Omega_{\rm vort} = \frac{16 \sqrt{2} \pi G \mu}{3 \sqrt{H_0}} \Omega_r^{3/4}  \frac{c v_b}{\zeta_b^3} P_{\rm vort} Y_b \sqrt{1-v_b^2} (1+Y_b) \left( \mathcal{I}(t_s) - \mathcal{I}(t_i) \right) ,
\end{equation}
where we define
\begin{equation}
    \mathcal{I}(t) = \int \frac{\text{e}^{-z_b t (t_0 - t)  }}{t^{3/2}}  d t
\end{equation}
with $z_b = \Gamma_d \alpha \zeta_b Y_b P_{\rm vort}$.

\section{Additional plots}
\label{D}

In this appendix, we include additional corner plots that demonstrate the influence of variations in MSCS network parameters analyzed with \texttt{PTArcade}. The corresponding plots are shown in Figs.~\ref{Fig.PTArcade_alpha}, \ref{Fig.PTArcade_F}, and \ref{Fig.PTArcade_e}.

\bibliographystyle{ourbst}
\bibliography{Bib}

\end{document}